\documentclass[showpacs,aps,floatfix,twocolumn,nofootinbib,preprintnumbers]{revtex4}

\usepackage{amssymb}
\usepackage{bm}
\usepackage{graphicx}

\begin{document}
\title{Shell--mediated tunnelling between (anti--)de Sitter vacua}
\author{Stefano Ansoldi}
\email{ansoldi@trieste.infn.it}
\homepage{http://www-dft.ts.infn.it/~ansoldi}
\affiliation{Center for Theoretical Physics -
Laboratory for Nuclear Science
and Department of Physics,
Massachusetts Institute of Technology\\
and International Center for Relativistic Astrophysics (ICRA) - Pescara - ITALY}
\altaffiliation{(\emph{Mailing address\/}) Dipartimento di Matematica e Informatica - Universit\`{a} degli Studi di Udine, via delle
Scienze 206, I-33100 Udine (UD), Italy}
\altaffiliation{INFN, Sezione di Trieste, Trieste, Italy}
\author{Lorenzo Sindoni}
\email{sindoni@sissa.it}
\affiliation{International School for Advanced Studies, SISSA/ISAS\\
via Beirut, 2-4 --- I-34014 Miramare, Trieste (TS), ITALY\\
and INFN, Sezione di Trieste}

\begin{abstract}
We give an extensive study of the tunnelling between arbitrary
(anti-)de Sitter spacetimes separated by an infinitesimally thin
relativistic shell in arbitrary spacetime dimensions. In particular,
we find analytically an exact expression for the tunnelling amplitude.
The detailed spacetime structures that can arise are discussed, together
with an effective \emph{regularization scheme} for \emph{before tunnelling}
configurations.
\end{abstract}


\preprint{MIT-CTP-3732}

\maketitle

\section{Introduction}

The study of vacuum decay initiated more than 30 years ago with the work of Callan and Coleman
\cite{bib:PhReD1977..15..2929C,bib:PhReD1977..16..1762C}; in the following
years the interest in the subject increased and the possible interplay of true vacuum
bubbles with gravitation was also studied \cite{bib:PhReD1980..21..3305L,bib:PhReD1987..36..1088W},
together with bubbles collisions and their importance in the early universe
\cite{bib:PhReD1982..26..2681S,bib:PhReD1983..28..1898C}.
At the same time, and as opposed with the true vacuum bubbles of Coleman \textit{et al.}, false
vacuum bubbles were also considered. In connection with gravity, the behavior of regions of false vacuum,
first studied by Sato et \textit{al.}
\cite{bib:PrThP1981..65..1443M,bib:PrThP1981..66..2052M,bib:PrThP1981..66..2287S,%
bib:PhLeB1982.108....98K,bib:PhLeB1982.108...103M,bib:PrThP1982..68..1979S}, was
for example analyzed in \cite{bib:PhReD1987..36..2919T,bib:PhReD1987..35..1747G,%
bib:PhReD1987..35..2961S,bib:NuPhy1990B339...417G,bib:PhReD1991..43.R3112T,%
bib:PhLeB1993.317...510Y,bib:NuPhB1999.556...373T,bib:PhReD2000..62083501T}.
For additional papers analyzing it in the context of inflation, we refer the reader to
\cite{bib:PhReD1983..27..2848V,bib:NuCiL1984..39...401L,bib:SPhyJ1984..60...211L};
interesting links with more phenomenologically oriented approaches can also be drawn,
as witnessed, for instance, by the recent \cite{bib:hepph2007..03...246M}.

In these last works, a minisuperspace approximation was adopted to quantize the
system. In more detail, since general relativistic shells\footnote{In this paper we will use interchangeably
the terms \emph{shell}, \emph{bubble} and \emph{brane}, which in different \emph{epochs} have appeared
in the literature to designate the system under consideration.} can be used as a convenient
model and since in spherical symmetry the system only has one degree of freedom,
standard semiclassical methods might be suitable to analyze the decay process
(see \cite{bib:PhReD1989..40..3982R} for an early, \emph{in principle}, discussion of
this point). In connection with cosmology, spaces \emph{equipped} with a cosmological
constant (\textit{i.e.\/} de Sitter space and generalizations) have been naturally considered.
In this context it also worth to remark the important role that they play in connection
with the problem of gravitational entropy, causal structure and the presence of horizons
(see, for instance \cite{bib:PhReD1977..15..2738H,bib:PhReD1995..52..5659H,bib:PhReD1996..54..6312H,%
bib:PhReD1996..53..7103G,bib:PhReD2000..61023502T} as well as
the suggestive \cite{bib:ClQuG1999..16...149D}).

Notwithstanding many interesting results, after 30 years, and with different flavors,
the problem of the stability of (the de Sitter) vacuum in connection with the dynamics
of false vacuum bubbles, is still a debated one
\cite{bib:NuPhB1999.556...373T,bib:ClQuG2005..22...283M,bib:PhReD2005..72103525J}.

The still open issues are highly non-trivial and go back to the, also long lasting, problem
of formulating a consistent framework for a quantum theory of gravity \cite{bib:PhReD1978..18..1747H,%
bib:PhReD1983..28..2960H,bib:PhReD1984..30...509V,bib:PhReD1992..46..4387C,%
bib:PhReD1998..58067301V,bib:PhReD1999..59084004C,bib:ClQuG1998..15..2605G}, but we will not
take explicitly this point of view here.
We will instead analyze a specific situation, described below, in which the
nucleation rate can be computed exactly in arbitrary spacetime dimensions. We will, then, explicitly
compute in closed form the nucleation rate in the semiclassical approximation, compare it with existing results,
and discuss in detail the associated spacetime structures: an analysis of how quantum effects may be
relevant in the context of
\emph{tunnelling from nothing configurations} \cite{bib:PhLeB1982.117....25V} will also be given.
These results, extend some results present in the literature (for instance \cite{bib:PhReD1990..41..2638P,%
bib:PhReD1990..42..4042P,bib:PhReD1994..49..6327G,bib:NuPhy1990B339...417G}) and can also
provide a useful limiting case of more general situations, for instance those discussed in
\cite{bib:PhReD2005..72103525J}.
At the same time, to consider spacetimes of higher dimensionality and negative cosmological constant is
especially important in view of recent results in the context of the AdS/CFT correspondence
\cite{bib:JHEP.2006..03...007S} and of the braneworld cosmological scenario
\cite{bib:JHEPh2000..10.....1O,bib:JHEP.2006..10...052K} (see also \cite{bib:PhReD2006..74126002K}
and references therein for a study in the context of noncommutative branes).

Apart from the papers already cited above, the instanton approach has also been discussed
by other authors (see for instance \cite{bib:PhReD1997..56..4651E,bib:PhReD1997..56..4663E,%
bib:ClQuG1997..14...L59M}) and although it will not be directly related to the present paper,
we cannot avoid mentioning the suggestive relationship between the decay of the cosmological
constant, membranes generated by higher rank gauge potentials and black holes, which
have appeared in many papers in the literature \cite{bib:NuPhy1980B176...509T,%
bib:PhLeB1984.143...415T,bib:PhLeB1984.147...258S,%
bib:NuPhB1985.252...523S,bib:PhLeB1987.195...177T,%
bib:NuPhy1988B297...787T,bib:ClQuG1989...6..1379M,%
bib:NuPhy2001B602...307W,bib:PhReD2001..64025008S,%
bib:PhReD2004..69083520W,bib:IJThP2004..43...883M}.

The structure of the paper is the following: in section~\ref{sec:thishe} we recall the formalism
by which a general relativistic thin spherical brane/shell can be described; this also gives
us the opportunity to fix conventions and notations and briefly describe the canonical approach to
its semiclassical quantization (subsection~\ref{subsec:canqua});
a convenient dimensionless formalism is also introduced (subsection~\ref{subsec:dimlesfor}).
We then directly come (section~\ref{sec:tuncal}) to the main result of this paper,
which is the calculation of the tunnelling rate between the classical configurations of the system,
in arbitrary spacetime dimensions; the results for the cases of $3$, $4$ and $5$ spacetime dimensions
are explicitly presented with dedicated plots of the values of the action as a function of the
two dimensionless parameters of the model (subsection~\ref{subsec:tuncalparcas}).
The exact results for the tunnelling amplitude/probability calculated in section~\ref{sec:tuncal}
correspond to specific transitions which take place in spacetime and that will be discussed
later on, in a dedicated appendix. We discuss, instead, in the main text
(section~\ref{sec:tunspatimstr}) a proposal to regularize some spacetime configurations which appear
to be singular, relating this issue to the description of the brane energy-matter content. In the concluding
section~\ref{sec:con}, the results of the paper are summarized; two appendices
follows with the detailed description of the parameter space of the system (appendix~\ref{app:parspaglostr})
and of all the Penrose diagrams associated with different values of the parameters of the problem
(appendix~\ref{app:clasol}): they are crucial to fully grasp the physical system under consideration.

\section{\label{sec:thishe}Shell in $N+1$ dimensions}

In this section we are going to briefly review some results
about the dynamics of co-dimension one branes in an $(N + 1)$-dimensional spacetime,
where, under the words \emph{co-dimension one brane}, we understand an $N$-dimensional
hypersurface $\Sigma$ separating the $(N + 1)$-dimensional manifold in
two domains, ${\mathcal{M}} _{-}$ and ${\mathcal{M}} _{+}$ having
$\Sigma$ as a common part of their boundary: in brief $\partial {\mathcal{M}} _{-}
\cap \partial {\mathcal{M}} _{+} = \Sigma$. In what follows we are
going to use the clear notation of \cite{bib:PhReD1991..43..1129I}:
the formulation developed there can be readily extended to higher
dimensions, just by letting the indices run on an extended set of values. We
will thus quickly report this paraphrase with the purpose of
recalling some notations and conventions. In particular let us
choose two arbitrary systems of $N+1$ independent vector fields
$\bm{E} _{\pm(a)}$ in ${\mathcal{M}} _{\pm}$, respectively, with
dual forms $\bm{\Omega} _{\pm} ^{(b)}$. Denoting with $\bm{g}
_{(\pm)}$ the four dimensional metric tensors in the two manifolds
we can write\footnote{In a few equations below, we will
use the notation $\langle - , - \rangle$ to denote the
scalar product. We also anticipate the notation
$\bm{\nabla} _{\bm{Y}} \bm{X}$ for the covariant derivative
of the vector fiend $\bm{X}$ in the direction of the
vector field $\bm{Y}$.}, in general,
\[
    \bm{g} _{(\pm)}
    =
    g _{(\pm) a b} \bm{\Omega} _{(\pm)} ^{(a)} \otimes \bm{\Omega} _{(\pm)} ^{(b)}
    ;
\]
in our notation Latin indices $a, b$ (as well as all other latin indices) will vary
in the set $\{ 0 , 1 , 2 , \dots{} , N \}$.

Let us first concentrate our attention on $\Sigma$: it is also a manifold, as
${\mathcal{M}} _{\pm}$, and we will denote by $\bm{e} _{(\mu)}$ an $N$-dimensional
system of (commuting) independent vector fields on $\Sigma$, with dual system $\bm{\omega} ^{(\nu)}$;
the indices $\mu$ and $\nu$ (as well as all other Greek indices) will vary in the set
$\{ 0 , 1 , \dots{} , N - 1 \}$.
Because of the physical interpretation we will be interested in the case in
which $\Sigma$ is embedded as a \emph{timelike}
surface in ${\mathcal{M}} _{\pm}$. Since $\Sigma$ is timelike the normal to $\Sigma$
in ${\mathcal{M}} _{\pm}$ is a spacelike vector $\bm{n}$, which we choose normalized so
that $\langle \bm{n} , \bm{n} \rangle = +1$. Moreover, our convention is that the normal
points from the ``$-$'' to the ``$+$'' domain of spacetime.

The components of $\bm{n}$ (an \emph{unambiguously} defined \emph{non-null} vector)
will be different, in general, when measured by an observer in ${\mathcal{M}} _{-}$
or by one in ${\mathcal{M}} _{+}$, according to the following definition:
\[
    n _{a} \vert _{\pm} = \langle \bm{n} , \bm{E} _{(a)} \rangle \vert _{\pm}
                        = \langle \bm{n} , \bm{E} _{\pm{}(a)} \rangle
    .
\]
In terms of $\bm{n}$ the \emph{extrinsic curvature} of the surface $\Sigma$
can be expressed as\footnote{We stress the normalization condition on $\bm{n}$.}
\[
    K_{\mu \nu}
    =
    \langle \textbf{n} , \bm{\nabla} _{\bm{e} _{(\mu)}} \bm{e} _{(\nu)} \rangle
\]
and, in general, it is different on the $-$ and $+$ side.
Moreover, by the Gauss-Codazzi
formalism, the geometry of spacetime around $\Sigma$ can be described in terms of the
intrinsic geometry of the hypersurface and of its extrinsic curvature. In the spirit of
this formalism, let us denote by $\bm{h}$ the intrinsic metric of the hypersurface $\Sigma$,
\textit{i.e.\/}
\[
    \bm{h} = h _{\mu \nu} \bm{\omega} ^{(\mu)} \otimes \bm{\omega} ^{(\nu)}
    .
\]
When we will work with quantities defined in the \emph{bulk} we will need to distinguish
the ones defined in ${\mathcal{M} _{+}}$ from the ones defined in ${\mathcal{M} _{-}}$,
and to this end we will use, as we did above, ``$\pm$'' superscripts or subscripts.
In many cases we will also be interested in the \emph{jump} of these quantities across
$\Sigma$: for instance, if we consider the extrinsic curvature, we may need the
to consider the difference $K ^{+} _{\alpha \beta} - K ^{-} _{\alpha \beta}$: following
\cite{bib:NuCim1966.B44.....1I,bib:PhReD1991..43..1129I}
we are going to rewrite this difference as $[K _{\alpha \beta}]$. Throughout this
paper this will be \emph{the only meaning} that we will give to the square brackets,
\textit{i.e.\/}
\[
    [A] \stackrel{\mathrm{def.}}{=} A ^{+} - A ^{-}
    .
\]
To avoid confusion no other use of the square brackets will be done.

\subsection{Junction conditions}

The brane $\Sigma$ can be more than just a mathematical surface,
\textit{i.e.\/} we can (and, in most of the cases,
we want to) equip it with a matter-energy content: it is then an
infinitesimally\footnote{Strictly speaking $\Sigma$ is
a source defined in a distributional way: for additional material on this point, the reader
is referred, for instance, to \cite{bib:PhReD1987..36..1017T}.} thin distribution of
matter-energy. Thanks to the above mentioned Gauss-Codazzi formalism, we can rewrite
Einstein equations to make explicit the contribution from the localized matter.
Then, the dynamics of $\Sigma$ as a surface separating ${\mathcal{M}} _{+}$ from
${\mathcal{M}} _{-}$ is obtained solving the following system of equations,
\begin{equation}
    K ^{+} _{\alpha \beta} - K ^{-} _{\alpha \beta}
    \equiv
    [ K _{\alpha \beta} ]
    =
    8 \pi G _{N+1} \left( S _{\alpha \beta} - h _{\alpha \beta} S / 2 \right)
    ;
\label{eq:juncon}
\end{equation}
these are Israel's junction conditions
\cite{bib:NuCim1966.B44.....1I,bib:PhReD1991..43..1129I}
which relate the jump in the extrinsic curvature
$[K _{\mu \nu}]$, \textit{i.e.\/} the ``jump'' in the way the surface is embedded in each geometry,
to the stress-energy tensor $S _{\mu \nu}$ of the matter contained
on $\Sigma$. The tensor $\bm{S}$ must also satisfy a conservation equation,
\[
    \langle {}^{(N)}{}\bm{\nabla} , \bm{S} \rangle _{\mu}
    =
    [ \bm{T} ( \bm{n} , \bm{e} _{(\mu)} ) ]
    ,
\]
where $\bm{T}$ is the stress energy tensor describing the content of
the complete spacetime manifold. Once we have specified the matter
content of the bulk (and hence the geometry according to Einstein
equations) the description of the dynamics of the system is
obtained by solving the system of equations (\ref{eq:juncon});
we refer the reader to \cite{bib:PhReD1991..43..1129I} and
\cite{maartensreview} for additional material and related
considerations.

\subsection{\label{subsec:sphsym}Spherical symmetry}

The set-up that is of interest for us is a simplified one, in which all the system
is spherically symmetric and the surface stress-energy tensor is that of the, so called,
\emph{tension model}, with
\[
    \bm{S} = - \kappa \bm{h}
    ,
\]
where $\kappa>0$ is a constant called the tension of the brane. In what follows,
it will be convenient to also define
\[
    \tilde{\kappa} = 4 \pi (N - 2) G_{N+1} \kappa .
\]
Thanks to the spherical symmetry, the \emph{system} of equations (\ref{eq:juncon}) can
then be reduced to a single equation \cite{ansoldi1},
\begin{equation}
    R
    \left(
        \epsilon _{-} \sqrt{\dot{R}^{2} + f _{-} (R) }
        -
        \epsilon _{+} \sqrt{\dot{R}^{2} + f _{+} (R) }
    \right)
    =
    \tilde{\kappa} R ^{2}
    ;
\label{eq:sphjuncon}
\end{equation}
in the above, $f _{\pm} (r _{\pm})$ are the metric functions of the static
line element adapted to the spherical symmetry, \textit{i.e.\/}, taking the
four dimensional case as a convenient example, we choose in both ${\mathcal{M}} _{\pm}$ the
coordinate system $x ^{a} _{\pm} = (t _{\pm} , \theta _{\pm} ,
\phi _{\pm} , r _{\pm})$ corresponding to the basis vectors
$\bm{e} _{t _{\pm}}$, $\bm{e} _{\theta _{\pm}}$, $\bm{e} _{\phi _{\pm}}$ and $\bm{e} _{r _{\pm}}$,
such that the metric is reduced to the form
$g _{(\pm) ab} = \mathrm{diag}
    (
    - f _{\pm} (r _{\pm}) ,
    r _{\pm} ^{2} ,
    r _{\pm} ^{2} \sin ^{2} \theta _{\pm} ,
    1 / f _{\pm} (r _{\pm})
    )
$. Generalizations to spacetimes with higher dimensionality are just
more cumbersome to write but trivial in their substance.
In these coordinate systems, we denote the radius of the brane
by $R ( \tau)$, where $\tau$ is the proper time of an
observer comoving with the brane and an overdot denotes the
derivative with respect to $\tau$. Moreover, as anticipated in the
introduction, we consider that this $N$-dimensional brane separates two
$(N + 1)$-dimensional spacetimes of the de Sitter/anti--de Sitter type, in
general with different cosmological constants $\Lambda _{\pm}$. We
thus have
\[
    f _{\pm} (r _{\pm}) = 1 - \frac{2 \Lambda _{\pm}}{N (N - 1)} r _{\pm} ^{2}
    .
\]
Finally $\epsilon _{\pm}$ are signs, defined as
\[
    \epsilon _{\pm} (R) = \mathrm{sign} ( \langle \bm{n} , \bm{e} _{r _{\pm}} \rangle ) | _{r _{\pm} = R}
\]
and are crucial quantities to obtain the Penrose diagrams associated
with the considered brane configuration.

\subsection{\label{subsec:canqua}The effective action/the momentum}

Before proceeding with the analysis of the physical system that
we introduced in the previous subsection, we believe it is useful
to recall some important points about the structure of the junction
conditions in spherical symmetry. For a generic junction between
spherically symmetric spacetimes across a (spherical) brane carrying
matter described by a given, but otherwise arbitrary,
equation of state, it would prove very useful to extract an effective
action for the dynamics of the brane starting directly from the
Einstein-Hilbert action and from a suitable action for the
matter fields and the brane \cite{ansoldi1}. This is the content
of a consistent literature on the subject \cite{bib:PhReD1997..56..4706B,%
bib:PhReD1997..56..7674W,bib:PhReD1998..57...914K,%
bib:PhReD1998..57...936H,bib:PhReD1998..57..2279W,bib:PhReD1998..58084005H,%
bib:PhReD2000..61901(E)K,2002PhReD..6502024028M,bib:ClQuG2004..21..5563R} among which we would like
to single out the more general, and recent, \cite{Barrabes:2005ar}:
the problem is not trivial at all, because of the same subtleties that appear, for
instance, in the Hamiltonian formulation of General Relativity.
Among the various possible approaches we will closely follow the quantization
procedure originally proposed by Farhi, Guth and Guven \cite{bib:NuPhy1990B339...417G},
which has, lately, been followed also in \cite{ansoldi1} and \cite{alberghi}.
For the Lagrangian approach to the classical dynamics we will instead
follow the clear and more general exposition of \cite{Barrabes:2005ar}.
We summarize here the relevant elements of these approaches referring the reader to
the literature for additional details. In particular, the effective action for the
shell can be obtained starting from the Einstein-Hilbert action with the
Gibbons-Hawking boundary terms \cite{bib:PhReD1977..15..2752H,bib:NuPhy1990B339...417G,Barrabes:2005ar}.
Using the Gauss-Codazzi formalism \cite{gravitation}, it is readily seen that
the Einstein-Hilbert action for a spherical co-dimension one brane (which in
our case will separate two domains of (anti--)de Sitter spacetime) can be decomposed
in non-dynamical bulk contributions (indeed, we consider the geometry of the
bulk spacetimes ${\mathcal{M}} _{\pm}$ fixed) and a
dynamical \emph{boundary} contribution described in terms of the extrinsic curvature
of the brane $\Sigma$. It is enlightening (and, perhaps, necessary) not to
use all the freedom in fixing the coordinate systems. As an exemplification
we do not use the freedom given by the reparametrization invariance with
respect to the proper time of the brane and we, thus, introduce a lapse
function $\mathcal{N} (s)$, so that the brane induced metric can be written as
$h ^{({\mathcal{N}})} _{\mu \nu} =$
$\mathrm{diag} ($
$- {\mathcal{N}} (s) ^{2} ,$
$R ^{2} (s) , \dots )$
in the coordinates $(s,R,\dots)$ (the ``\dots{}'' stand for the
trivial spherically symmetric part). Then, following \cite{ansoldi1,Barrabes:2005ar},
the effective action for the degrees of freedom associated with the radial
and time coordinates takes the form
\begin{widetext}
\begin{equation}
    S_{\mathrm{eff}}
    \propto
    \int
       \left\{
       R ^{N-2}
          \left[
            \epsilon \sqrt{\dot{R} ^{2} + {\mathcal{N}} ^{2} f (R)}
            -
            \dot{R} \,
            \mathrm{arctanh} \,
            \left(\frac{\dot{R}}{\epsilon \sqrt{\dot{R} ^{2} + {\mathcal{N}} ^{2} f (R)}} \right) ^{\mathrm{sign}(f)}
          \right]
       - \tilde{\kappa} {\mathcal{N}} R^{N-1}
    \right\} d\tau .
\label{eq:effaction}
\end{equation}
Thus, the effective Lagrangian of the system is given by
\[
    {\mathcal{L}} = P ^{({\mathcal{N}})} \dot{R} - {\mathcal{H}}
    ,
\]
where, following \cite{ansoldi1}, we have defined the effective Hamiltonian as
\[
    {\mathcal{H}} = - R^{N-2}\left[\epsilon \sqrt{\dot{R} ^{2} + {\mathcal{N}} ^{2} f (R)}\right] + \tilde{\kappa} \, {\mathcal{N}}R^{N-1}
\]
and the effective momentum as
\begin{equation}
    P ^{({\mathcal{N}})} (R , \dot{R}) =
    - R ^{N-2}
    \left[
       \mathrm{arctanh}
       \left(
           \frac{\dot{R}}{\epsilon \sqrt{\dot{R} ^{2} + {\mathcal{N}} ^{2} f (R)}}
       \right) ^{\mathrm{sign}(f)}
    \right]
    .
\label{eq:conjmom}
\end{equation}
\end{widetext}
Then, the equations of motion for the $R$ and $\mathcal{N}$ degrees of freedom are
\[
    \frac{d\mathcal{H}}{d\tau}= \frac{\dot{{\mathcal{N}}}}{{\mathcal{N}}} {\mathcal{H}}
    \quad
    \mathrm{and}
    \quad
    {\mathcal{H}}(R,\dot{R},{\mathcal{N}})=0;
\]
the second equation is a first integral of the former one, and is the
\emph{Hamiltonian constraint}. We see
that it encodes all the information about the dynamics of the system,
being identical to the only remaining junction condition. The first equation
is the second order equation of motion of the system, given by the total
derivative with respect to the proper time of the Hamiltonian constraint.
In what follows, we will be mostly interested in the expression of the
momentum evaluated on a solution of the equation of motion.
In order to build a canonical structure (as it is done, for
example, in \cite{bib:JHEPh2000..10.....1O}) we first have to prove the existence of a
well defined symplectic structure on the phase space of the Lagrangian system;
in particular the Legendre transform has to be invertible, which
in our case means nothing but the invertibility of the conjugate momentum
as a function of the velocity \cite{arnold}. As appreciated already
in \cite{bib:NuPhy1990B339...417G}, for generic co-dimension one branes this cannot
be proved to be always satisfied\footnote{In these cases the quantum theory cannot be built
with the canonical formalism. Alternative formulations using the
path integral approach have been considered in the seminal work
of Farhi, Guth and Guven \cite{bib:NuPhy1990B339...417G}.}. However,
for the system under consideration the canonical
structure always exists and $P ^{({\mathcal{N})}}$ can rigorously be considered
the canonical momentum conjugate to the coordinate $R$; thus, in the
specific case of interest here, the canonical construction of the
corresponding quantum theory is a well posed problem.
To prove this statement, let us consider
\[
\frac{\partial P ^{({\mathcal{N}})}}{\partial \dot{R}} = - R^{N-2} \left[
\frac{1}{\epsilon \sqrt{\dot{R} ^{2} + {\mathcal{N}}^{2} f(R)}} \right]
.
\]
This quantity can be zero
if and only if two conditions are
simultaneously satisfied:
\begin{enumerate}
  \item $\epsilon_{+}=\epsilon_{-}$;
  \item there is a point $R$ in the configuration space such that
$f_{+}(R)=f_{-}(R)$.
\end{enumerate}
For our particular choice of branes in (anti--)de Sitter spacetime,
we see that in order
for the second condition to hold we must have $\Lambda_{-}=\Lambda_{+}$;
but this implies\footnote{This
can be easily seen from the junction condition
(\protect\ref{eq:sphjuncon}) remembering that $\tilde{\kappa} > 0$.}
$\epsilon_{+}=-1, \epsilon_{-}=+1$
so that the first condition then fails.
Thus, for the class of junctions that we are considering
one can always build a canonical structure.

\subsection{\label{subsec:dimlesfor}Dimensionless formalism}

Following, for instance, \cite{bib:PhReD1989..40..2511S},
it is possible to determine the classical dynamics
of the system by studying an equivalent one dimensional problem for
the radial degree of freedom $R$, taking also into account the values of the signs
$\epsilon _{\pm} (R)$.

Before proceeding we will set up a different system of dimensionless quantities,
in order to remove the arbitrariness in the definition of the normalization of the
cosmological constants and of the bubble's tension. If we define the quantities
\begin{eqnarray}
    & &
    x
    =
    \tilde{\kappa} R ,
    \nonumber \\
    & &
    \bar{\tau}
    =
    \tilde{\kappa} \tau ,
    \nonumber \\
    & &
    \lambda _{\pm}
    =
    \frac{2 \Lambda _{\pm}}{\tilde{\kappa} ^{2} N (N-1)} ,
    \nonumber \\
    \nonumber \\
    & &
    \qquad \qquad \alpha = \lambda_{-} + \lambda_{+} ,
    \nonumber \\
    & &
    \qquad \qquad \beta = \lambda _{-} - \lambda_{+} ,
    \nonumber
\end{eqnarray}
equation (\ref{eq:sphjuncon}) then becomes\footnote{An
overdot will denote, from now on, a derivative with respect to $\bar{\tau}$.}
\begin{equation}
    - \left[ \epsilon \sqrt{\dot{x}^2 + 1-\lambda x^2} \right] x = x^2
    .
    \label{eq:adijuncon}
\end{equation}
For completeness we also report the dimensionless form of the effective momentum
(\ref{eq:conjmom}) when we impose the gauge choice ${\mathcal{N}} \equiv 1$:
\begin{equation}
    \bar{P} (x , \dot{x}) = - x ^{N-2}
    \left[
       \epsilon \; \mathrm{arctanh}
       \left(
           \frac{\dot{x}}{\sqrt{\dot{x} ^{2} + f (x)}}
       \right) ^{\frac{f}{|f|}}
    \right]
    .
\label{eq:adiconjmom}
\end{equation}
This quantity will be central in what follows, but for the moment we keep our
attention on equation (\ref{eq:adijuncon}).
Despite its unusual look, it is easily proved that it is equivalent to a
system of equations, which, in the notation that we are using, takes the form
\begin{equation}
    \left\{
        \matrix{
            \dot{x} ^{2} + \bar{V} (x) = 0 \hfill
            \cr
            \epsilon _{\pm} (x) = - \mathrm{sgn} ( \beta \pm 1 ) \hfill
        }
    \right .
    ,
\label{eq:claequequ}
\end{equation}
where the potential $\bar{V} (x)$ has a simple parabolic form
\begin{equation}
    \bar{V} (x) = 1 - \frac{x ^{2}}{x _{0} ^{2}}
\label{eq:effpot}
\end{equation}
and $x _{0}$, the turning point, is given by
\begin{equation}
     x _{0} ^{2} = \frac{4}{{1 + 2 \alpha + \beta^2}}
    ,
\label{eq:turpoi}
\end{equation}
when the right-hand side is positive; otherwise there are no nontrivial
classical solutions. All the solutions of the problem can be easily
classified in according to the values of $\alpha$ and $\beta$: we refer
the reader to appendix~\ref{app:parspaglostr} for details. The expressions for
the signs $\epsilon _{\pm}$ relate, instead, the global geometrical structure
of spacetime for this brane configuration only to the difference
between the inner and outer cosmological constants, not to the
details of the trajectory itself. This is a portion of the content of the
junction condition, that is necessary for the description of the
global spacetime structure: a careful discussion of this point can
be found in appendix~\ref{app:clasol}.

The form of equation (\ref{eq:adijuncon}) for a brane separating two
(anti-)de~Sitter spacetimes, which, in turn, is responsible for the
simple quadratic form of the potential (\ref{eq:effpot}), allows exact
solution of this particular case. It is, in fact,
not difficult to see that (\ref{eq:adijuncon}) has:
\begin{enumerate}
    \item the \emph{trivial solution} $x (\bar{\tau}) \equiv 0$, which always exists;
    \item the solution
        \begin{equation}
            x (\bar{\tau})
            =
            x _{0} \cosh \left( \frac{\bar{\tau} - \bar{\tau} ^{(0)}}{x _{0}} \right)
            ,
        \label{eq:clasol}
        \end{equation}
        which satisfies the initial condition $x (\bar{\tau} ^{(0)}) = x _{0}$;
        this solution is known as the \emph{bounce solution} and exists only if the condition
        \[
            {1 + 2 \alpha + \beta ^{2}} > 0
        \]
        is satisfied.
\end{enumerate}
We also, incidentally, note that
\[
    \beta = \pm 1 \: \Longleftrightarrow \: \frac{1}{x _{0} ^{2}}
    =
    \lambda_{\mp}
\]
\textit{i.e.\/} when $|\beta|=1$ the turning point radius coincides
with one of the two cosmological horizons, if they exist.
We also anticipate that a more careful analysis of the
trivial solution $x (\bar{\tau}) \equiv 0$
will be required, since, although
it is given to us by the mathematics of the problem, we
are mostly interested in its physical role, especially
when we will turn on (semiclassical) quantum effects
(see section \ref{sec:tunspatimstr}).

\section{\label{sec:tuncal}Tunnelling}

Using the notation introduced in the last section
we can now describe how the semiclassical regime of the brane
dynamics looks like. In particular we can consider the
tunnelling between the zero-radius solution towards the bouncing
solution (\ref{eq:clasol}), and vice versa.

For example in one direction the semiclassical picture is as follows: we have a brane, of very small radius;
of course, when it is very small, the quantum properties of its matter content will be non-negligible
(we will further discuss this issue later on in section \ref{sec:tunspatimstr}) and their interplay
with gravity will be non-trivial; although
we do not know the full quantum gravity description of the system, we will consider that,
thanks to quantum effects, the brane will have a certain probability to tunnel under the potential
barrier given by the effective potential (\ref{eq:effpot}) into the bounce solution.
When emerging after the tunnelling quantum effects
will likely become less and less important as compared with gravitational
ones (and as far as the interaction with the bulk spacetime is concerned),
so that the evolution of the brane will closely resemble that of the
classical junction. This is not the case in the first stage of the evolution,
involving the tunnelling process, and we will try to understand, at least at an effective
level, the ``responsibilities'' of both the quantum and gravitational realms in this
process. In particular, quantum effects will be considered in a modification
at small scales of the stress-energy tensor (see section \ref{sec:tunspatimstr}).
Gravitational effects, at small scales, are also present,
and described by the geometric character of Israel junction conditions. We think
that our description might be able to take into account both these effect,
at least within the limits represented by the semiclassical approximation.
In this sense, although our treatment will be effective,
it will give us the possibility to consistently solve the ambiguity
arising at the mathematical level and represented by the $x (\bar{\tau}) \equiv 0$
solution. We will suggest that, at the semiclassical level, a more detailed treatment
of both, the quantum and gravitational aspects, is unlikely to be necessary.

\subsection{Tunnelling trajectories}

For the system under consideration, tunnelling trajectory can be
described by constructing the instanton in the Euclidean sector.
The construction of the instanton in the general case is still an
unsolved problem: we will follow the approach of \cite{bib:NuPhy1990B339...417G};
moreover, the problems
left open in \cite{bib:NuPhy1990B339...417G} do not affect the present case.
In fact it is easy to see that in our case the instanton describing
the tunnelling can always be constructed without the necessity to
introduce the \emph{pseudo-manifold} that in \cite{bib:NuPhy1990B339...417G}
was necessary to deal with multiple covering of points in the
Euclidean sector. Moreover the different descriptions of the tunnelling process
that were obtained in \cite{bib:NuPhy1990B339...417G} using the canonical or
the path-integral approaches in our case also coincide, because they
are consequences of properties of the dynamical variables during the tunnelling
trajectory which are not present in the system that we are considering
(please, see \cite{Ansoldi:2004hb,Ansoldi:2007xu,Ansoldi:2007hb} for a more detailed
description of these issues).

In this way we know that the tunnelling can be described directly
at the effective level, where, the relevant aspects of the Euclidean
junction can be determined by the Wick rotated classical effective
system. We thus define $\bar{\tau} _{\mathrm{e}} = - i \bar{\tau}$;
then, denoting with a prime the derivative with respect to
$\bar{\tau} _{\mathrm{e}}$, we have
that the Euclidean system is obtained with the formal substitutions
$
\dot{x} \rightarrow i x'
$
for the ``velocity'' and
\[
\bar{P} \rightarrow i \bar{P} _{\mathrm{e}}
\]
for the momentum (it is not difficult to prove that the above
\emph{substitution rules} are rigorous results and that they can be
obtained performing an \emph{Euclidean junction}, as done, for instance,
in \cite{bib:NuPhy1990B339...417G}). In particular
\begin{equation}
\bar{P}_{\mathrm{e}} (x , x') = - x^{N-2} \left[
\arctan\left(\frac{x'}{\epsilon \sqrt{f (x) - (x')^2}} \right) \right]
.
\label{eq:equeffconmom}
\end{equation}
In this way, the tunnelling process can be modelled using
the effective tunnelling trajectory, which solves the Euclidean
equation:
\begin{equation}
- (x') ^{2} + 1 - (x/x_0)^2 = 0 .
\label{eq:euceffequ}
\end{equation}
The analytic expression of the tunnelling trajectory of a brane
expanding from zero radius at Euclidean time $\bar{\tau} _{e} = 0$ to the bouncing solution is:
\[
x(\bar{\tau}_{\mathrm{e}}) = x_0 \sin(\bar{\tau}_{\mathrm{e}}/ x_0) .
\]
The opposite process is then described by
\[
x(\bar{\tau}_{\mathrm{e}}) = x_0 \cos(\bar{\tau}_{\mathrm{e}} / x_0)
,
\]
if we assume that at Euclidean time $\bar{\tau} _{\mathrm{e}} = 0$ the brane starts contracting
from $x _{0}$.
These two processes occur with certain probabilities, whose amplitudes
$\mathcal{A}$ can be expressed in the usual semiclassical approximation
as
\[
\mathcal{A}(0\rightarrow x_0) \propto \exp(-I_{\mathrm{e}}[x])
,
\]
where $I _{\mathrm{e}}[x]$ is the Euclidean and can be
obtained as the integral of the Euclidean momentum evaluated on
a solution of the Euclidean equation of motion (\ref{eq:euceffequ}).
If, for simplicity, we define
\[
I_{\mathrm{e}}[x] = \frac{\Omega_{N-1}}{16 \pi G_{N+1}} \frac{1}{\kappa^{N-1}}
\bar{I}_{\mathrm{e}}[x],
\]
where $\Omega_{N-1}$ is the volume of ${\mathbb{S}}^{N-1}$,
then $\bar{I}_{\mathrm{e}}[x]$ can be obtained as
\begin{equation}
\label{eq:pidx} {\bar{I}} _{\mathrm{e}} [x] = \int_{0}^{x_0} \bar{P} _{\mathrm{e}} (x)
dx
.
\end{equation}
We stress again that $\bar{P} _{\mathrm{e}} (x)$ is the Euclidean momentum evaluated on a
solution of the Euclidean equation of  motion and can be obtained by substituting
$x ' = \sqrt{\bar{V} (x)}$ in (\ref{eq:equeffconmom}). We thus have
\[
    \bar{P} _{\mathrm{e}} (x)
    =
    -
    x ^{N-2}
    \left[
        \arctan \left( \frac{2 \sqrt{\bar{V} (x)}}{(\beta + \omega) x} \right)
    \right]
    ,
\]
where, to make writing more compact\footnote{In view of the definition of the $\omega$'s,
and of the meaning of the square brackets, the equation above is a shorthand for
$
    \bar{P} _{\mathrm{e}} (x)
    =
    -
    x ^{N-2}
    \left(
        \epsilon _{+}
        \arctan \left( \frac{2 \sqrt{\bar{V} (x)}}{(\beta + 1) x} \right)
        -
        \epsilon _{-}
        \arctan \left( \frac{2 \sqrt{\bar{V} (x)}}{(\beta - 1) x} \right)
    \right)
$.}, we have introduced the quantitie\emph{s}
$\omega _{\pm}$ defined as $\omega _{\pm} = \pm 1$.

Note that in the Euclidean case we have some freedom in appropriately
choosing one of the possible branches of the inverse tangent
function. Different choices will affect, in general, the result that we
obtain for the action. As in \cite{bib:NuPhy1990B339...417G} we
observe that non-careful choices will make the action a discontinuous
function of the parameters; this can be seen without difficulties.
Preliminarily, let us anticipate that with the symbol ``$\arctan$'',
we will  indicate the branch of the inverse tangent function with range
in $[-\pi/2,\pi/2]$. Let us then consider $\beta \neq \omega$ and
let us integrate by parts the integral (\ref{eq:pidx}). We obtain
\begin{eqnarray}
    {\bar{I}} _{\mathrm{e}} [x]
    & = &
    -
    \left . \left [
        \frac{x ^{N-1}}{N-1} \arctan \left( \frac{2 \sqrt{\bar{V} (x)}}{(\beta + \omega) x} \right)
    \right ] \right | ^{x _{0}} _{0}
    +
    \nonumber \\
    & & \qquad +
        \left [
            \int _{0} ^{x _{0}} \frac{x ^{N-1}}{N-1} d \left( \arctan \left( \frac{2 \sqrt{\bar{V} (x)}}{(\beta + \omega) x} \right) \right)
        \right ]
    .
    \nonumber
\end{eqnarray}
Let us now consider the first term above. Clearly, if we chose the branch of the $\arctan$
function to be the one with range in $[-\pi/2,\pi/2]$, then we have
\[
    \lim _{\beta \to - \omega ^{-}}
        \left . \left [
            \frac{x ^{N-1}}{N-1} \arctan \left( \frac{2 \sqrt{\bar{V} (x)}}{(\beta + \omega) x} \right)
        \right ] \right | ^{x _{0}} _{0}
    =
    -
    \frac{\omega \pi x _{0} ^{N-1}}{2 (N - 1)}
\]
but
\[
    \lim _{\beta \to - \omega ^{+}}
        \left . \left [
            \frac{x ^{N-1}}{N-1} \arctan \left( \frac{2 \sqrt{\bar{V} (x)}}{(\beta + \omega) x} \right)
        \right ] \right | ^{x _{0}} _{0}
    =
    +
    \frac{\omega \pi x _{0} ^{N-1}}{2 (N - 1)}
    ,
\]
so that the action develops a discontinuity at $\beta = - \omega$. This discontinuity can be eliminated
if we choose different branches of the inverse tangent functions (please remember that in the
expression that we are considering we are using a compact notation to indicate the difference of
two inverse tangent functions); in particular the two discontinuities at $\beta = \pm 1$
can be eliminated by choosing:
\begin{enumerate}
    \item the branch with range $[- \pi , 0]$ for the $\arctan$ function containing
quantities of the ``$+$'' spacetime;
    \item the branch with range $[- \pi , 0]$ for the $\arctan$ function containing
quantities of the ``$-$'' spacetime.
\end{enumerate}
Since in our notation ``$\arctan$'' has range $[-\pi/2,\pi/2]$,
this means that in the equations above must be rewritten with the substitutions
($\Theta$ is the step function)
\[
    \arctan (\mbox{``}+\mbox{''}) \longrightarrow \arctan (\mbox{``}+\mbox{''}) - \pi \Theta (\beta + 1 )
\]
and
\[
    \arctan (\mbox{``}-\mbox{''}) \longrightarrow \arctan (\mbox{``}-\mbox{''}) - \pi \Theta (\beta - 1 )
    ;
\]
for the \emph{jump} of the quantity which appears inside the expression of
the Euclidean momentum, this implies the following substitution:
\[
    [ \arctan ] \longrightarrow [ \arctan ] - \pi \Theta ( 1 - \beta ^{2} )
    .
\]
In this way the action integral is continuous also at the points $\beta = - \omega$
and is given by the integral
\begin{eqnarray}
    & &
    {\bar{I}} _{\mathrm{e}} [x]
    =
    -
    \int_{0}^{x_0} dx
    x ^{N-2} \times
    \nonumber \\
    & & \quad
    \times
    \left(
    \left[
        \arctan \left( \frac{2 \sqrt{\bar{V} (x)}}{(\beta + \omega) x} \right)
    \right]
        - \pi \Theta ( 1 - \beta ^{2} )
    \right)
    .
\label{eq:pidxcon}
\end{eqnarray}
After these preliminary considerations, we can proceed to evaluate the
tunnelling amplitude in the WKB approximation; as we anticipated and as
we will see, in arbitrary spacetime dimensions this result can be
expressed analytically in terms of known functions.

\subsection{\label{subsec:gentunres}General result for the tunnelling amplitude}

To calculate the first contribution to the integral (\ref{eq:pidxcon}) we
proceed as follows. As a preliminary step, we observe that it can be written as a difference
of two integrals with the same general structure.
Let us then consider the two integrals containing the inverse tangent functions:
small differences, which
do not substantially affect the calculation, can be taken into account by
properly using the $\omega$'s introduced above, as we already did
in the expressions for the momentum. This said, we can perform\footnote{Strictly
speaking this can be done only when $\beta \neq - \omega$. The cases in which
$\beta = - \omega$ are trivial and can be dealt separately; or, more simply,
since we already know from the discussion in the previous subsection that the
action is continuous, we can just extend it to $\beta = - \omega$ by continuity.}
an integration by parts in (\ref{eq:pidxcon}). The terms evaluated at the
limits of integration, which appear in this process, do vanish and by changing the
integration variable from $x$ to $\zeta = (x / x _{0}) ^{2}$ (which also transforms the
integration domain into the unit interval $(0,1)$) we obtain the following expression,
\begin{eqnarray}
&&
{{\bar{I}}} _{\mathrm{e}} [x] = - \frac{x_0^{N}}{4(N-1)}\times
\nonumber \\
&& \quad \times
\left[(\beta+\omega) \int_0 ^1
\frac{\zeta^{(N-2)/2}}{\sqrt{1-\zeta}}\frac{1}{1-z_{\omega}\zeta}
d\zeta \right]+
\label{eq:actintold} \\
&& \qquad
+
\frac{\pi x _{0} ^{N-1}}{(N-1)}
    \Theta (1 - \beta ^{2})
,
\nonumber
\end{eqnarray}
where
\begin{equation}
\label{eq:pole} z_{\omega} =
\left(1-x_0\frac{\beta+\omega}{2}\right)\left(1+x_0\frac{\beta+\omega}{2}
\right)
.
\end{equation}
When $z _{\omega} < 1$ one of the above integrals diverges since there is
a pole of the integrand on the domain of integration. On the other
hand, it is easy to see that the condition $z _{\omega} < 1$ implies
\[
    \frac{x _{0} ^{2}}{4} ( \beta + \omega ) ^{2} < 0
\]
and cannot be realized for any choice of the parameters if a tunnelling
trajectory has to exist. The value
$z _{\omega}=1$ can instead be obtained if $\beta=-\omega$ and we know that
the action can be extended by continuity to these values, although the above procedure
to calculate the integral is not valid. Thus, under the conditions $\beta \neq - \omega$,
we have that $z _{\omega} > 1$ is satisfied and equation (\ref{eq:actintold}) gives
\begin{eqnarray}
& &
{{\bar{I}}} _{\mathrm{e}} [x] = -
\frac{x_0^{N}}{4(N-1)}
\frac{\Gamma(N/2)\Gamma(1/2)}{\Gamma((N+1)/2)} \times
\nonumber \\
& & \quad
\times \left[
(\beta+\omega){}_2 F _{1}\left( 1,
\frac{N}{2},\frac{N+1}{2},z_{\omega} \right) \right]
+
\label{eq:actintnew} \\
& &  \qquad +
\frac{\pi x _{0} ^{N-1}}{(N-1)} \Theta (1 - \beta ^{2})
,
\nonumber
\end{eqnarray}
where $\Gamma$ is the Euler's gamma function, ${}_{2}F _{1}$
the hypergeometric function. Note that $x_0$
depends on $\alpha,\beta$ and so the first factor of the formula,
depending on the number of spacetime dimensions, cannot be ignored
even for a qualitative description of the tunnelling amplitude.

\subsection{\label{subsec:tuncalparcas}Some cases of interest}

\begin{figure*}[!ht]
\begin{center}
\fbox{\includegraphics[width=17.6cm]{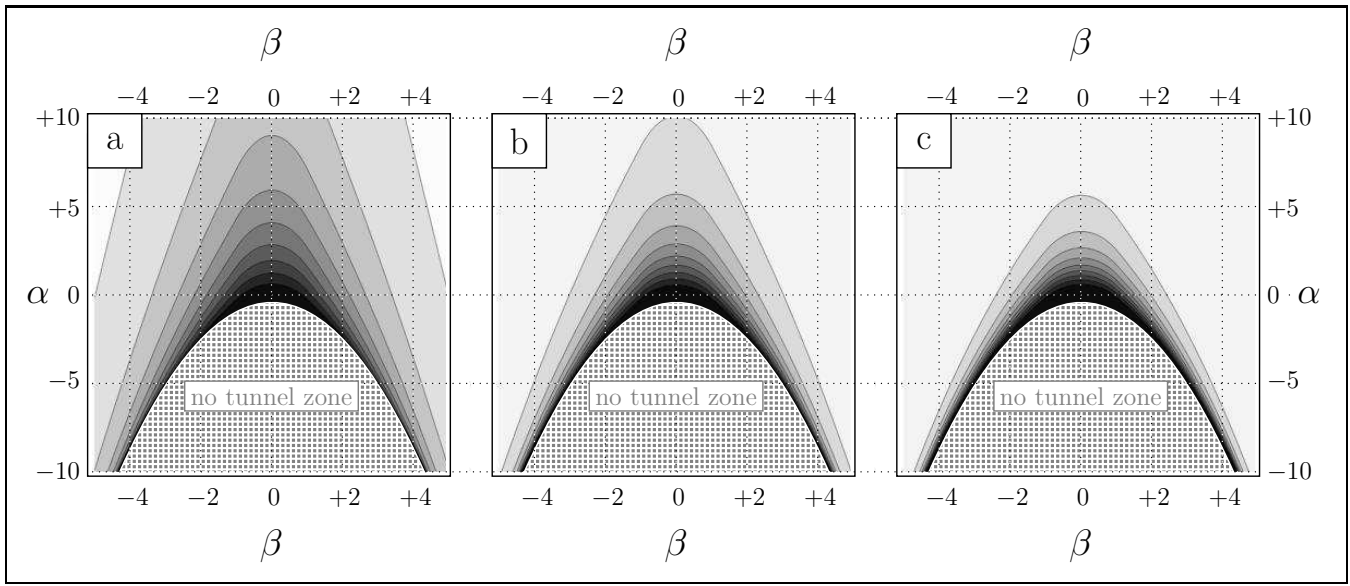}}
\end{center}
\caption{\label{fig:ActConPlo}Contour plots of the values of the \emph{tunnelling probability} in
$3$ (case a), $4$ (case b) and $5$ (case c) spacetime dimensions in the $\beta$--$\alpha$ plane.
The area inside the crosshatched parabola in the bottom of each picture is the region in parameter
space where tunnelling cannot occur since no bounce solutions exists (no tunnel zone).
A better understanding of what happens close to the
limiting parabola can be understood from the following figures~{\protect\ref{fig:3_Dact_02}},
{\protect\ref{fig:4_Dact_02}} and {\protect\ref{fig:5_Dact_02}}. Next to next contours indicate
a jump of the probability of $0.1$, with the lightest gray tone indicating the range $(0.9,1.0)$ and
the darkest the range $(0.0,0.1)$. The same color indicates the same range of values
of the probability in all the three sub-diagrams.}
\end{figure*}

It is useful to specialize the result (\ref{eq:actintnew}) to particular
situations. We are going to do this by considering the cases
in which spacetime is three, four and five dimensional.
Below we are going to explicitly discuss these three cases. A comparative
presentation of the results can be found in the contour plots of
figure~\ref{fig:ActConPlo}.

\begin{figure*}[!ht]
\begin{center}
\fbox{\includegraphics[width=12.8cm]{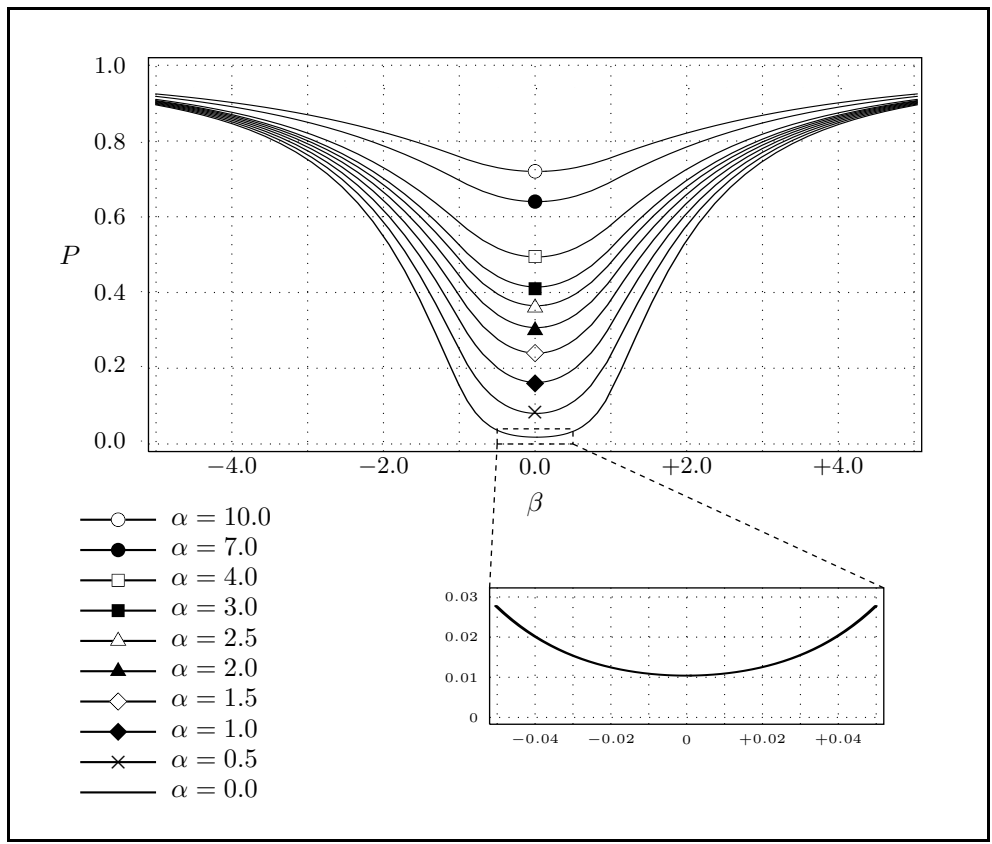}}
\end{center}
\caption{\label{fig:3_Dact_01}Plots of the values of the \emph{tunnelling probability} in a
spacetime of dimension $3$,
as a function of $\beta$ for fixed non-negative values of $\alpha$ as listed above. The detailed behavior
for $\alpha = 0$ around $\beta = 0$, is also shown (to better see that the probability is
small but non-zero).}
\end{figure*}

\begin{figure*}[!ht]
\begin{center}
\fbox{\includegraphics[width=12.8cm]{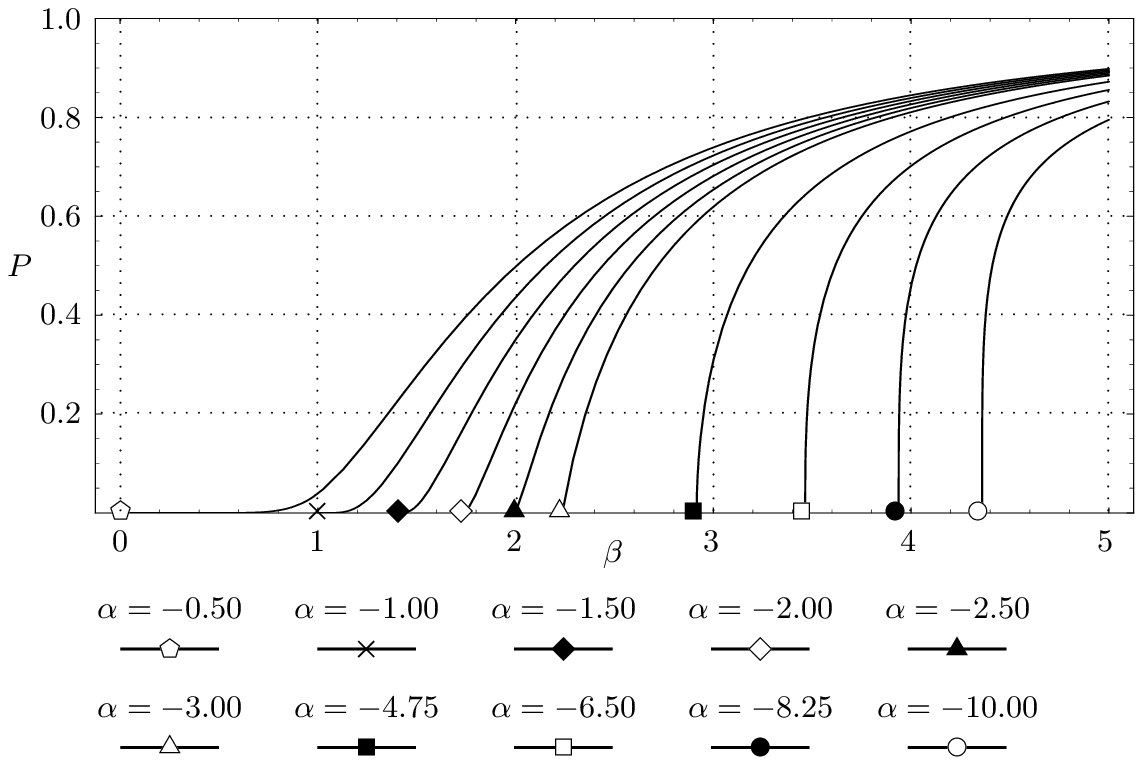}}
\end{center}
\caption{\label{fig:3_Dact_02}Plots of the values of the \emph{tunnelling probability} in a
spacetime of dimension $3$, as a function of $\beta$ for fixed negative values of $\alpha$ as listed
above. For $\alpha < - 1/2$, not all values of $\beta$ are allowed. We have put the mark on each
curve in correspondence of the minimum value of $\beta$ for which (at the given value of $\alpha$)
the tunnelling process can exist. For smaller values, the infinitely expanding solution, in fact,
is not present.}
\end{figure*}

We start then with the three dimensional case, which is lower-dimensional gravity; in three
spacetime dimensions it seems more clear how to build a quantum theory out of the
classical junction conditions \cite{bib:JHEPh2000..10.....1O}. Moreover three dimensional
gravity is an interesting system by itself, it allows an easier visualization
of some results and can be used for interesting specific toy models. Anyway, in this case
we have $N = 2$ and we can express $_{2} F _{1}\left( 1,1;3/2;x\right)$ in terms of
elementary functions as
\begin{equation}\label{eq:hyper3d}
    {}_2F{}_1\left( 1,1;\frac{3}{2};y\right) =
\frac{\arcsin(\sqrt{y})}{\sqrt{1-y}\sqrt{y}}
.
\end{equation}
Correspondingly the action becomes
\begin{eqnarray}
&&
    {\bar{I}} _{\mathrm{e}}
    =
    -
    \frac{x _{0} ^{2}}{4}
    \frac{\Gamma (1) \Gamma (1/2)}{\Gamma(3/2)} \times
\nonumber \\
&&
\quad \times
\left[ (\beta+\omega)
    \frac{\arcsin(\sqrt{z_{\omega}})}{\sqrt{1-z_{\omega}}\sqrt{z_{\omega}}}
    \right]+
\label{eq:act3d} \\
&& \qquad \vphantom{\frac{\arcsin(\sqrt{z_{\omega}})}{\sqrt{1-z_{\omega}}\sqrt{z_{\omega}}}}
+
\pi x _{0} \Theta (1 - \beta ^{2})
.
\nonumber
\end{eqnarray}
The corresponding probability is plotted as a function of $\beta$ in figure \ref{fig:3_Dact_01}
for some non-negative values of $\alpha$ and in figure \ref{fig:3_Dact_02}
for some negative values of $\alpha$.

Not many comments are necessary, of course, for the four dimensional case,
the original arena on which this calculation was performed. Again we can take
advantage of a simple expression for the corresponding hypergeometric function
\begin{equation}\label{eq:hyper4d}
    {}_2F{}_1\left(1,3/2;2;y\right) = \frac{2}{y\sqrt{1-y}} - \frac{2}{y}
,
\end{equation}
which brings the final result in the form
\begin{eqnarray}
    & &
    {\bar{I}} _{\mathrm{e}}
    =
    - \frac{\pi x _{0} ^{2}}{4} \times
    \label{eq:act4d} \\
    & & \times
    \left(
        \left[
            \mathrm{sgn(\beta + \omega)}
            \frac{1-(1 - z _{\omega}) ^{1/2}}{z _{\omega}}
        \right]
    -
    \Theta (1 - \beta ^{2} )
    \right) .
\nonumber
\end{eqnarray}
\begin{figure*}[!ht]
\begin{center}
\fbox{\includegraphics[width=12.8cm]{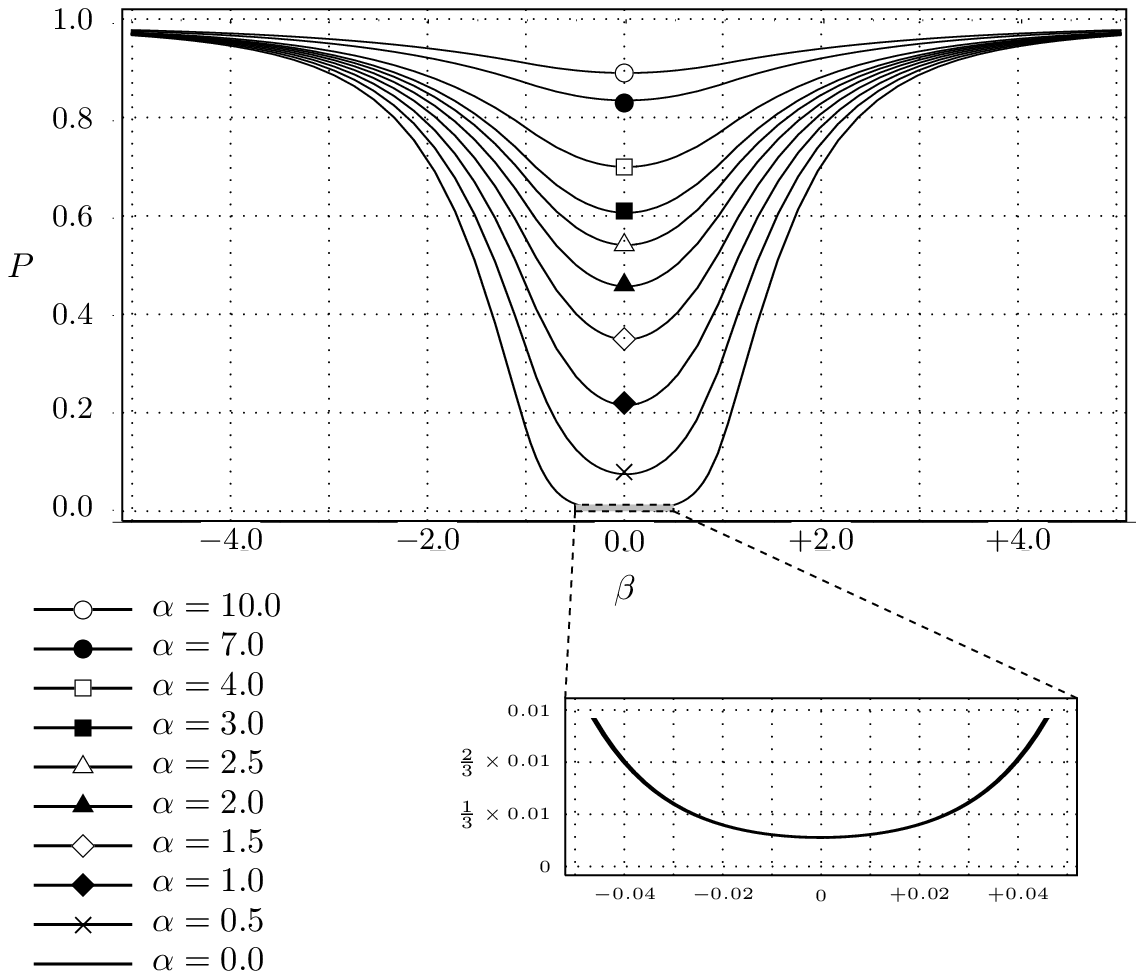}}
\end{center}
\caption{\label{fig:4_Dact_01}Plots of the values of the \emph{tunnelling probability} in a
spacetime of dimension $4$, as a function of $\beta$ for fixed non-negative values of $\alpha$ as listed above.
Again the detailed behavior for $\alpha = 0$ around $\beta = 0$ is shown (to better see that the probability,
although very small, is non-vanishing).}
\end{figure*}
\begin{figure*}[!ht]
\begin{center}
\fbox{\includegraphics[width=12.8cm]{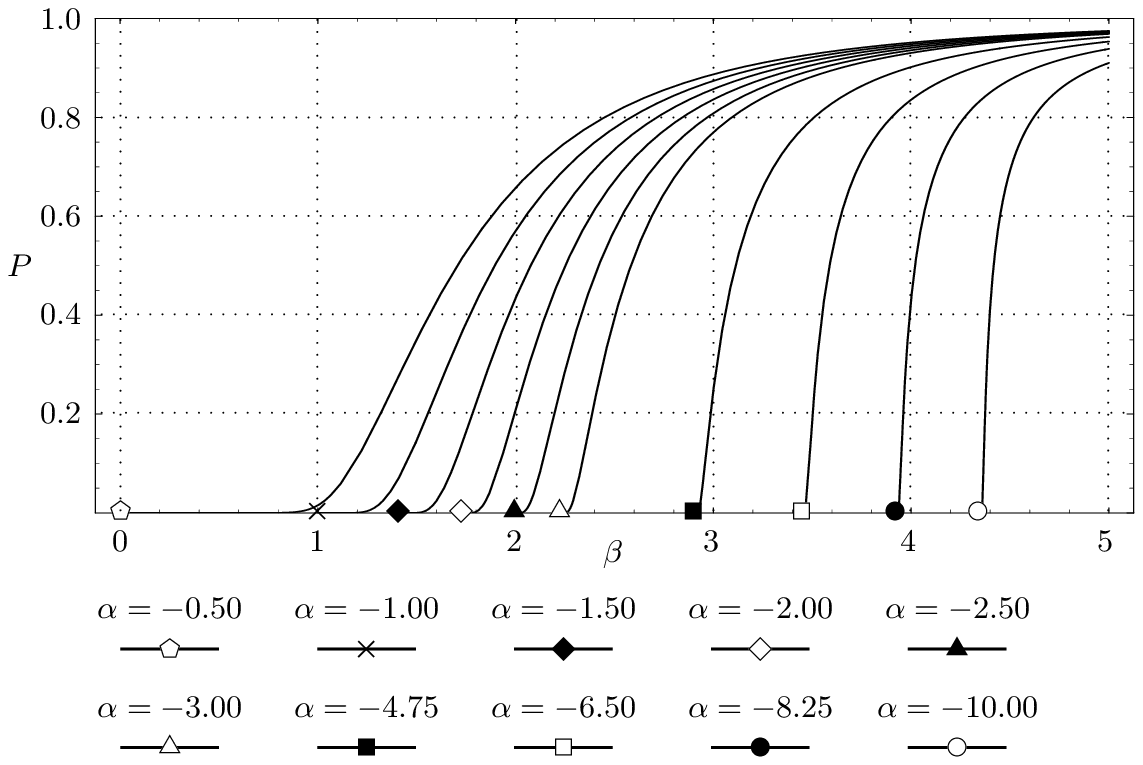}}
\end{center}
\caption{\label{fig:4_Dact_02}Plots of the values of the \emph{tunnelling probability} in a
spacetime of dimension $4$, as a function of $\beta$ for fixed negative values of $\alpha$ as listed
above. The same observations as in the case of $3$ spacetime dimensions apply.}
\end{figure*}
The plots of the probability as a function of $\beta$
can be found in figure \ref{fig:4_Dact_01} for some non-negative
values of $\alpha$ and in figure \ref{fig:4_Dact_02} for some negative
values of $\alpha$.  This result is the same as the one obtained
in \cite{bib:PhLeB1983.121...313P} and it also reduces to the one calculated
in \cite{bib:PhReD1980..21..3305L}: it, thus, represents a useful consistency check.

Finally, the five dimensional case is interesting in the context of
the Randall-Sundrum scenario. In this case, using
\begin{equation}\label{eq:hyper5d}
    _2F_1\left( 1,
2;\frac{5}{2};y\right) =
\frac{3}{2 y} \left(\frac{\arcsin(\sqrt{y})}{\sqrt{1-y}y^{1/2}} - 1\right)
,
\end{equation}
the result for the action integral can be put in the following form:
\begin{eqnarray}
& &
   {\bar{I}} _{\mathrm{e}} = - \frac{x_0^4}{12} \frac{\Gamma (2) \Gamma (1/2)}{\Gamma (5/2)}
\times
\nonumber \\
& & \quad \times
\left[ (\beta+\omega) \left(
  \frac{\arcsin(\sqrt{z_{\omega}})}{\sqrt{1-z_{\omega}}(z_{\omega})^{3/2}}
-\frac{1}{z_{\omega}}
  \right)
  \right]
+
\label{eq:act5d}
\\
& & \qquad
+ \frac{\pi x _{0} ^{3}}{3} \Theta (1 - \beta ^{2})
.
\nonumber
\end{eqnarray}
\begin{figure*}[!ht]
\begin{center}
\fbox{\includegraphics[width=12.8cm]{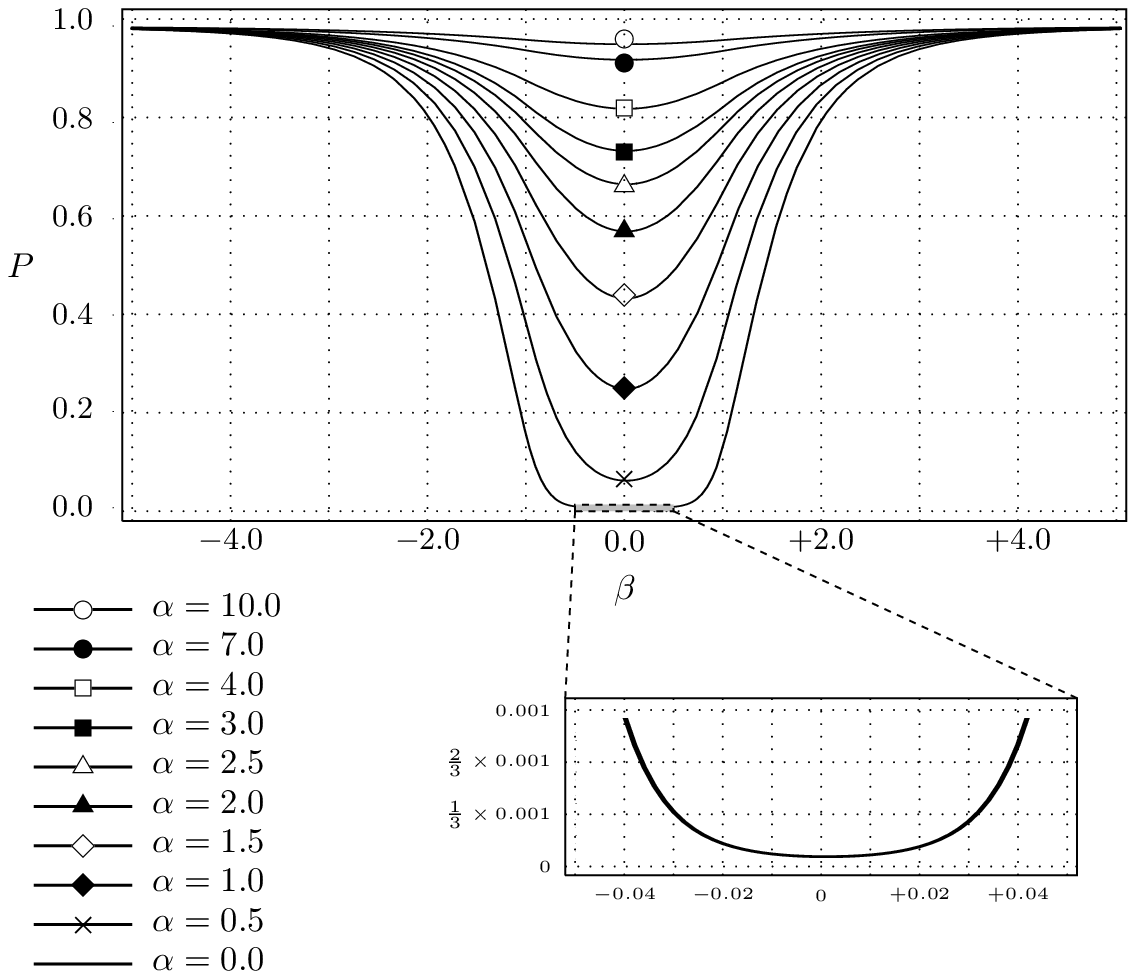}}
\end{center}
\caption{\label{fig:5_Dact_01}Plots of the values of the \emph{tunnelling probability} in a
spacetime of dimension $5$, as a function of $\beta$ for fixed non-negative values of $\alpha$ as listed above.
Again the detailed behavior for $\alpha = 0$ around $\beta = 0$ is shown (to better see that the probability,
although much smaller than in the previous cases, is still non-vanishing).}
\end{figure*}
\begin{figure*}[!ht]
\begin{center}
\fbox{\includegraphics[width=12.8cm]{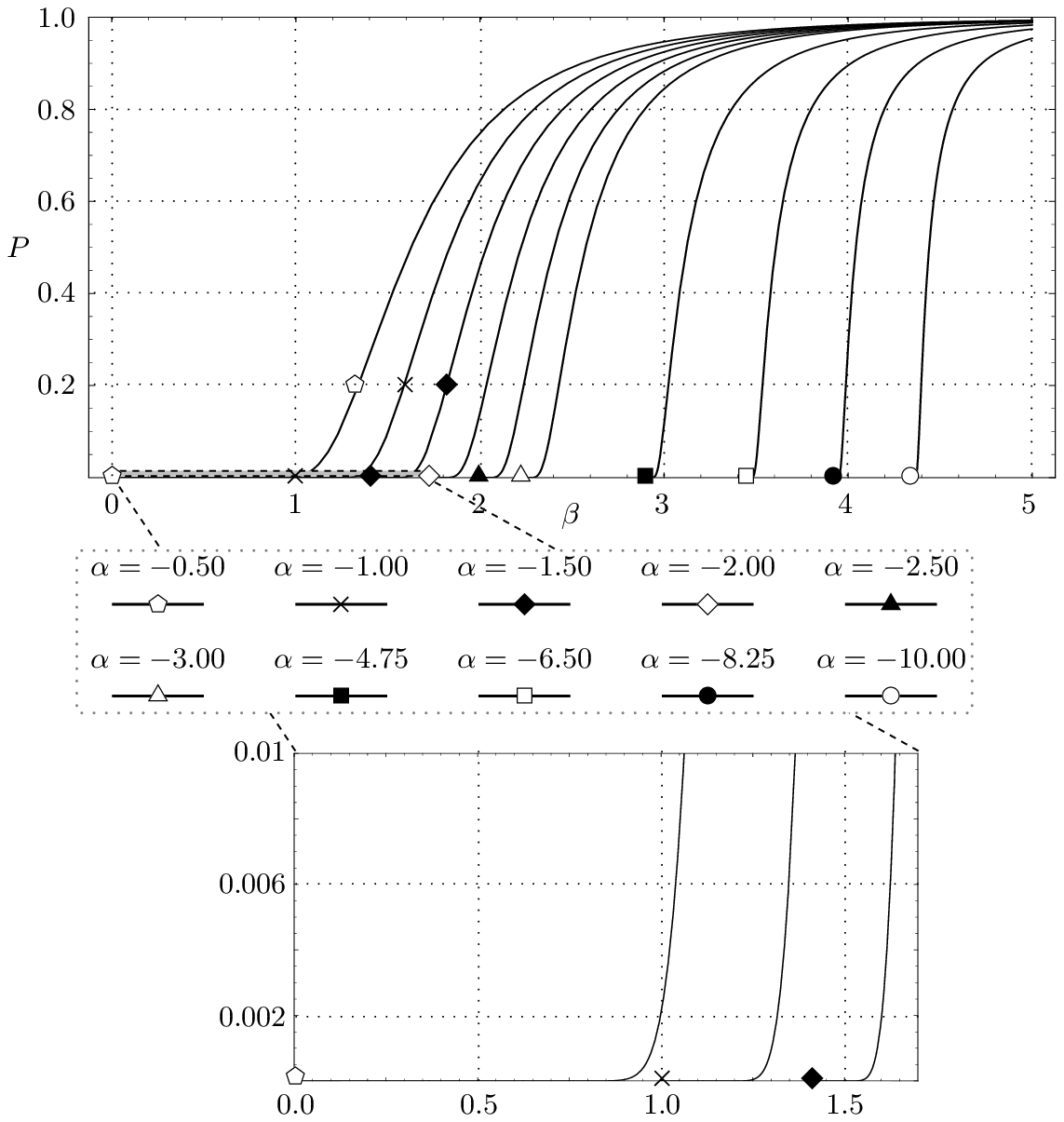}}
\end{center}
\caption{\label{fig:5_Dact_02}Plots of the values of the \emph{tunnelling probability} in a
spacetime of dimension $5$, as a function of $\beta$ for fixed negative values of $\alpha$ as listed
above. For $\alpha < - 1/2$, not all values of $\beta$ are allowed. We have put the mark on each
curve in correspondence of the minimum value of $\beta$ for which (at the given value of $\alpha$)
the tunnelling process can exist and, the zoomed diagram, helps resolving the superposition that
arises between the curves obtained for $\alpha=-0.50, -1.00, -1.50$.}
\end{figure*}
Again we present two plots of the corresponding probability as a function
of $\beta$; figure \ref{fig:5_Dact_01} shows the behavior for some non-negative
values of $\alpha$, whereas plots for some negative values of $\alpha$ can be
found in figure \ref{fig:5_Dact_02}.

\section{\label{sec:tunspatimstr}Discussion}

Up to this point we have discussed the physics of the system rather quickly,
focusing mainly on the mathematics necessary to describe the classical and the semi-classical phases.
There is still a point which deserves a detailed discussion, since it involves non-trivial aspects of
the dynamics of the brane. Indeed, the physical process we have considered is the tunnelling of spacetime,
from a classical situation \emph{representing} a spacetime containing a small brane to another
classical configuration \emph{describing} a bouncing brane.
In our picture, the pre-tunnelling state is represented by the $x \equiv 0$ solution of the junction
condition. Although this case might
appear rather simple at first sight and one could be tempted to just state that
the initial state is, for example, the full ${\mathcal{M}} _{+}$ spacetime
(a point of view which has been taken for example in \cite{ansoldi1}), much more care has to
be taken. The main reason is that when $x=0$, the brane's world-volume degenerates to a curve.
Thus, the formalism that we used to describe the brane breaks down and the analysis that we have
made cannot be considered as rigorous as it is in the case of the bouncing solutions.
As a manifestation of this fundamental problem, we observe that the equations determining the
$\epsilon _{\pm}$ signs do not hold for the particular solution $x \equiv 0$. In
particular, for this solution the junction condition as written in (\ref{eq:adijuncon})
does not provide any mean to solve this problem.

A way out of this situation appears when we reflect on the fact that this process,
although described semiclassically, is quantum in its true nature.
Thus an approximation of the system which
considers it completely classical \emph{before the tunnelling}, i.e. in the degenerate configuration,
might be too rough and might require a more careful consideration. During the tunnelling trajectory,
and even more for the $x \equiv 0$ classical
solution, quantum effects are supposed to be, if not dominant, at least relevant
enough to modify the classical picture of the junction.

We will propose here an attempt to address the problem. Our proposal should be considered
a first step further, but
far from a first principle solution of the complex quantum problem;
it just aims to show that the mathematical and physical aspects of the problem stated above
can be dealt with by giving an effective formulation for the phenomena that might arise
at small scales. At the same time, although we will just build an effective model, we
will not be very demanding about its main properties: in this way, hopefully, the model,
although effective, could mimic well enough effects produced by quantum gravity whatever
will be their, still undiscovered, true nature.

In this respect, we would also like to point out the following:
i) the main effect of our proposal is to \emph{slightly perturb} the effective
potential (\ref{eq:effpot}) in the tunnelling region; then, we will ii) show that the perturbed problem is free
from ambiguities and iii) use for the \emph{unperturbed case} the results
obtained in the \emph{perturbed one}, when the perturbation becomes smaller and smaller.
To have a definite model, we will regard the small
scale behavior of the matter composing the brane/shell as the physical origin for
the perturbations (see below). On the other hand, the consequence of
these perturbations in the effective formulation is completely \emph{generic}: it is,
in fact, possible to show that other physical motivations, as for instance the quantum
properties of spacetime at small scales, could be reflected in a similar way in the
effective formulation.
For these reasons, we can regard our conclusions \emph{model independent} to a high degree,
at least within the context defined by the semiclassical approximation.

To introduce our model we can naturally think the brane as sourced by
some matter fields whose nature is, ultimately, quantum. We thus argue that,
although at large scales the approximation for the brane stress energy tensor that we made in
subsection \ref{subsec:sphsym} might be rough but still appropriate, at small scales it
will instead break down due to quantum effects. We will model these additional quantum effects by
adding a term to the brane stress-energy tensor as follows:
\begin{equation}
    \bm{S} \longrightarrow \bm{S} + \bm{S} _{\mathrm{q}}
    ,
\end{equation}
where the ``quantum'' contribution $\bm{S} _{\mathrm{q}}$ will be pa\-ra\-me\-tri\-zed as
\begin{equation}
    \bm{S} _{\mathrm{q}} = \rho _{\mathrm{q}} \bm{u} \otimes \bm{u} + \sigma _{\mathrm{q}} \bm{h}
    .
\end{equation}
The conservation equation for $\bm{S} _{\mathrm{q}}$ under the assumption of spherical
symmetry gives (we are going to use, from now on, the dimensionless versions of the
parameters $\rho _{\mathrm{q}}$ and $\sigma _{\mathrm{q}}$, which following the notation above,
are called $\bar{\rho} _{\mathrm{q}}$ and $\bar{\sigma} _{\mathrm{q}}$)
\begin{equation}
    \frac{d \bar{\rho} _{\mathrm{q}}}{d x}
    +
    N
    \frac{\bar{\rho} _{\mathrm{q}}}{x}
    =
    \frac{d \bar{\sigma} _{\mathrm{q}}}{d x}
    .
\end{equation}
We will choose preliminarily
\begin{equation}
    \bar{\rho} _{\mathrm{q}} ( x ) = a x ^{q}
\end{equation}
so that
\begin{equation}
    \bar{\sigma} _{\mathrm{q}} (x) = a \frac{q + N}{q} x ^{q}
    .
\end{equation}
In this case the junction condition becomes\footnote{We remember that we are following
the convention in which square brackets indicate the jump of the enclosed quantity, \textit{i.e.\/}
the difference of it when evaluated in the `$+$' and `$-$' domains. We are now using the suffix
``$\dots{}_{\mathrm{q}}$'' or ``$\dots{}_{(\mathrm{q})}$''
to indicate quantities when the energy momentum tensor is modified, as described in the text, to take into account
$\mathrm{q}$uantum effects; thus the two signs will now be $\epsilon _{(\mathrm{q})\pm}$.}
\begin{equation}
    -
    \left[ \epsilon _{(\mathrm{q})} \sqrt{\dot{x} ^{2} + 1 - \lambda x ^{2}} \right] x = x ^{2} \mu (x)
    ,
    \label{eq:modadijuncon}
\end{equation}
with
\begin{equation}
    \mu (x)
    =
    1 + \bar{a} x ^{q}
\end{equation}
and, for short, $\bar{a} = a (q + N) / q$. Now we are going to slightly
restrict the parameters $\bar{a}$ and $q$ to make these general settings appropriate
for our model. In particular,
the modification to the stress energy tensor, described by $\bar{a}$ and $q$
was introduced to model the quantum effects at small scales; thus it should be
negligible at large scales and this can be achieved if $q + 1 < 0$, \textit{i.e.\/} $q < -1$.

We can now compute the modified effective potential $\bar{V} _{\mathrm{q}}$.
The potential turns out to be
\begin{eqnarray}
    \! \! \! \! \! \! \! \! \!
    \bar{V} _{\mathrm{q}} (x)
    & = &
    -
    \frac{x ^{2} \left[ \beta ^{2} + 2 \alpha (1 + \bar{a} x ^{q}) ^{2} + (1 + \bar{a} x ^{q}) ^{4} \right]}
         {4 (1 + \bar{a} x ^{q}) ^{2}}
    \label{eq:modeffpot} \\
    & \stackrel{x \rightsquigarrow 0}{\rightsquigarrow} &
    - \frac{\bar{a} ^{2}}{4 x ^{- 2q - 2}}
    \label{eq:modeffpotx_0} \\
    & \stackrel{\bar{a} \rightsquigarrow 0}{\rightsquigarrow} &
    \bar{V} (x) + \frac{(1 - \beta ^{2}) \bar{a} ^{2}}{2 x ^{- q - 2}}
    \label{eq:modeffpotd_0}
    .
\end{eqnarray}
Extracting the behavior of (\ref{eq:modeffpot}) for small $x$ we get (\ref{eq:modeffpotx_0}),
whereas for small $\bar{a}$ the leading contribution is given by (\ref{eq:modeffpotd_0}).
These are useful results. Indeed we see from (\ref{eq:modeffpotx_0}) that, quite generally,
and certainly under the above $q < -1$ condition, the potential satisfies
\[
    \lim _{x \to 0 ^{+}} \bar{V} _{\mathrm{q}} (x) = - \infty
    .
\]
This implies that the $\bar{V} _{\mathrm{q}} (x)$ allows not only the bounce brane junction,
as $\bar{V} (x)$ does, but also bounded solutions. Thus the addition of the $\bm{S} _{\mathrm{q}}$
term to the stress energy tensor, which in our picture is supposed to take into account quantum
effects at small scales, in fact does his job by trading the $x \equiv 0$ solution for a
bounded one of finite (\textit{i.e.\/} non-vanishing) size. Moreover this solution exists under
very general assumptions about the form of $\bm{S} _{\mathrm{q}}$, so that we do not
have to commit ourselves too much about the underlying quantum gravity physics of
which $\bm{S} _{\mathrm{q}}$ is roughly supposed to take into account some effective
semiclassical description. Looking, now at equation (\ref{eq:modeffpotd_0}) we also see
that for small $\bar{a}$, apart from the evident qualitative difference at small scales,
the potential resembles closely the non-perturbed one (and this happens, in particular,
along the tunnelling trajectory). Thus it will be a sufficiently good approximation
to evaluate the tunnelling probability in the, analytically much simpler, unperturbed case.

At this point, our picture of the spacetime transition will be as follows. We will
take as final configurations of the tunnelling process the junctions obtained without considering the
correction, which in our setup is strongly suppressed at large distances; then, the initial state, which is the
$x \equiv 0$ solutions, will be ``regularized'' by considering the junction as the
limiting case of the bounded perturbed junction when $\bar{a} \to 0$. In particular,
the sign ambiguity, which affects the $x \equiv 0$ solution, will be solved by
choosing the signs of the bounded trajectory of the perturbed model.
These signs can be obtained in closed form considering the $x \rightarrow 0$ limit
of the junction condition, which is dominated by the correction. Therefore
\begin{equation}
    \epsilon _{\pm}
     =\mp 1
\label{eq:modsig}
\end{equation}

Using this prescription, then, one can build the spacetime diagrams representing
the tunnelling process. We refer the reader to appendices for a complete
discussion of the parameter space (appendix~\ref{app:parspaglostr}) and of
the global spacetime diagrams representing the various physical situations
(appendix~\ref{app:clasol}).

To conclude the discussion, we would like to remark that, despite the fact that the regularization
procedure we have described changes dramatically the situation in the region near $x=0$ of the
configuration space of the brane, the effect on the calculation we have performed is not significant. This
is due to the fact that in the tunnelling region, the potential is substantially modified only in a narrow
region close to the turning point corresponding to the maximum radius of the bounded solution
of the modified classical junction condition. In the tunnelling region, the perturbation to the potential
is finite, and by sending $\bar{a} \rightarrow 0$ the tunnelling amplitude of the modified problem
is approximated arbitrarily well by our calculation in section~\ref{sec:tuncal}.
Of course, in the real physical problem we expect that quantum effects would prevent the brane
from shrinking to zero radius: thus the mathematical limit $\bar{a} \rightarrow 0$ should be read
as a more physically sound limit in which $\bar{a}$ tends to a small but finite value related to
the ultimate physical nature of the brane itself (which is presently only partially understood).
As a consequence, the analytical result that we have obtained (which relies on a
WKB approximated description of the quantum process) can also be consider an approximation
of the full quantum result at the lowest power in the ratio between the maximum radius of the bounded solution
of the modified classical junction equations and $x _{0}$.

\section{\label{sec:con}Conclusions}

In this paper we have calculated analytically the  tunnelling
amplitude for a domain of spacetime of de Sitter/anti--de Sitter type in a
background which, again, is de Sitter/anti--de Sitter. The analytical result holds in arbitrary spacetime dimensions greater than three,
and generalizes already existing four dimensional calculations. It is not difficult
to see that this result reduces, for appropriate values of the parameters, to the
result found by Coleman and de Luccia \cite{bib:PhReD1980..21..3305L} for the false
vacuum to true vacuum transition in four spacetime dimensions. Also the results
by Parke \cite{bib:PhLeB1983.121...313P}, again in the four dimensional case,
are correctly reproduced.

We have, also, discussed extensively, in the text and in appendix \ref{app:clasol},
the spacetime structures that can arise for all possible values of the parameters
characterizing the model. Some of these spacetime structures, already discussed in the
literature, are known as \emph{tunnelling from nothing} configurations. We have also exposed
a possible issue of the shell formalism in the description of the pre-tunnelling state and
proposed a solution which relies on the above mentioned \emph{tunnelling from nothing} configurations;
in this way we have been able to make what we consider to be a consistent choice
for the \emph{before tunnelling} configurations. This proposal,
which is implemented by considering a modification to the stress-energy tensor for the matter
on the shell at small states, is motivated by the observation that if quantum effects are
non-negligible on scales at which the tunnelling process occurs, they should also be non-negligible
at smaller scales, where the \emph{before tunnelling} configurations live. By modelling the influence
of these quantum effects with a quite generic modification to the form of the stress-energy tensor
at small scales, we have proposed an unambiguous rule to fix the initial configuration. This
approach to the problem, seems to us consistent with the level at which we are modelling quantum
effects for this gravitational system, which is the \emph{semiclassical approximation} for an
\emph{infinitesimally thin} distribution of matter and energy.
At the same time, it has already been shown that, if we add
a more refined matter content on the shell (as for instance
a collection of gauge fields), modifications similar to
the one that we have considered in this paper naturally appear
\cite{bib:ClQuG1999..16..3315P}.
Moreover, we note that it gives
a very consistent picture of the tunnelling process, since all the possible types of tunnelling
result in a ``sudden expansion'' of a very small region of spacetime from a small size (where quantum
effects are certainly non negligible) to a much bigger size, with radius of the order of $x _{0}$.
This shows that the regularized tunnelling always models what in the literature has been called
tunnelling ``from nothing''. In our case the ``nothing'' is exactly the quantum state of the
spacetime junction that we model, in an effective way, using (\ref{eq:modadijuncon}). It thus
seems that some tunnelling configurations present in the literature and of more
difficult interpretation \cite{ansoldi1} might be ruled out by the quantum properties of
matter and/or of spacetime at small scales. In fact, it is suggestive to reflect about the fact that,
even in our very simplified and purely effective treatment of quantum effects before and during the
tunnelling, all tunnelling processes start exactly ``from nothing'' (as interpreted above), i.e. from
a configuration which is consistent with the quantum properties of the following tunnelling process.

To conclude, we would like to explicitly distinguish between the analytical result for the tunnelling
probability and the proposal to interpret the \emph{pre-tunnelling} configurations. Indeed
the analytical result is a natural generalization of already existing calculations and it incorporates
them as special cases.  Its validity is completely independent from our interpretation of the
\emph{before tunnelling} configurations and it can represent a useful limit case to check the results of more
elaborated models: for instance, a de Sitter--Schwarzschild shell configuration should reproduce, in the limit of
vanishing Schwarzschild mass, our result for a tunnelling between a de Sitter space of assigned cosmological
constant and a de Sitter space with vanishing cosmological constant, i.e. Minkowski space.

\begin{acknowledgments}
We would like to thank M. C. Johnson and S. Sonego for useful discussions and constructive
criticism.
This work is supported in part by funds provided by the U.S. Department of
Energy (D.O.E.) under cooperative research agreement \#{}DF-FC02-94ER40818.
The work of S. Ansoldi is supported in part by a grant from the Fulbright Commission.
\end{acknowledgments}

\appendix

\section{\label{app:parspaglostr}Parameter space}

In this appendix we will elaborate about the classification of the possible junctions
between two (anti--)de Sitter spacetimes. After switching to the dimensionless formulation
(see subsection~\ref{subsec:dimlesfor}),
we remain with two parameters, \textit{i.e.\/} $\lambda _{\pm}$, or,
which is the same, $\alpha$ and $\beta$. We will mostly use
the latter quantities in the following analysis, since many
relevant features of the solutions to Israel's junction
condition related to the causal structure of the full
spacetime manifold ${\mathcal{M}}$ are easily deducible from
the comparison between the two cosmological constants.
\begin{figure}
\begin{center}
\includegraphics[width=8.5cm]{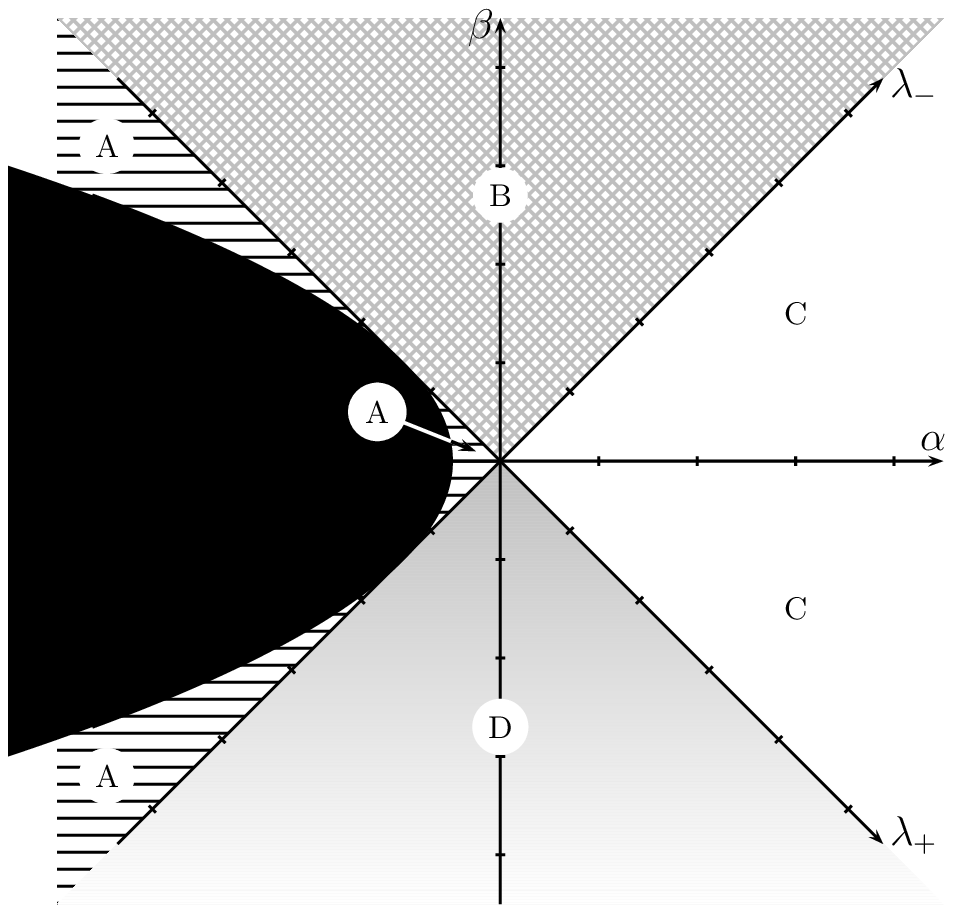}
\end{center}
\caption{\label{fig:parspa}The space of parameters is subdivided according
to the different kinds of junctions, which can be between anti--de Sitter and
anti--de Sitter spacetimes (type A), anti--de Sitter and de Sitter spacetimes
(type B), de Sitter and anti--de Sitter spacetimes
(type D) or de Sitter and de Sitter spacetimes (type C). The black part
of the parameter space, identified by the condition $1 + 2 \alpha + \beta ^{2} < 0$,
singles out the values of the parameters for which tunnelling is not possible since
the infinitely expanding solution does not exist.}
\end{figure}
In particular, in figure \ref{fig:parspa} we give a classification
of the possible solutions. The diagram shows the parameter space
of the variables $(\alpha , \beta)$ \emph{and} $(\lambda _{+} , \lambda _{-})$.
The classification of the solutions using $\alpha$ and $\beta$, looks
\emph{nicely symmetric} with respect to the axis $\beta = 0$,
for which $\lambda _{+} = \lambda _{-}$; this is the primary reason
why we will mostly use the $(\alpha , \beta)$ parametrization in
our discussion. The $(\lambda _{+} , \lambda _{-})$ axes can, anyway,
be conveniently used to single out the de Sitter from the anti--de Sitter
spacetime.
In particular, we have defined four main groups of solutions:
\begin{description}
    \item[type A)] these are solutions in which both spacetimes
    joined across the brane are anti--de Sitter spacetimes, i.e we have an
    $\mathrm{AdS} _{(-)}-\mathrm{AdS} _{(+)}$ junction; this part of the parameter
    space is bounded by the parabola $\alpha = - (\beta ^{2} + 1) / 2$
    (equivalently $x _{0} ^{-1} = 0$), whose inside is the black region
    where no solution exists;
    \item[type B)] these solutions describe a junction of one part of de Sitter
    spacetime in the ${\mathcal{M}} _{-}$ manifold with a part of anti--de
    Sitter spacetime in the ${\mathcal{M}} _{+}$ manifold, \textit{i.e.\/} they are
    $\mathrm{dS} _{(-)}-\mathrm{AdS} _{(+)}$ junctions, and play the role of counterparts of the
    below discussed type D solutions;
    \item[type C)] these are junctions in which both spacetimes have the
    de Sitter geometry, \textit{i.e.\/} we have $\mathrm{dS} _{(-)}-\mathrm{dS}_{(+)}$ junctions; in some
    cases, qualitatively different diagrams may arise depending on which
    cosmological constant is the bigger, \textit{i.e.\/} depending on the sign of $\beta$;
    \item[type D)] as anticipated above, this last type of solutions is similar
    to the type B, with the role of ``$-$'' and ``$+$'' interchanged; we thus
    have $\mathrm{AdS} _{(-)}-\mathrm{dS}_{(+)}$ junctions.
\end{description}
The above information is not enough to completely characterize the classical
spacetime obtained from the junction. In addition we need to know the behavior
of the normal to the brane travelling in the spacetimes that we have determined
from the diagram in figure (\ref{fig:parspa}).
According to our convention the normal to the brane
has its tail-tip direction going from the ``$-$'' to the ``$+$'' parts of
the full spacetime ${\mathcal{M}}$; on the other hand in each of the two spacetimes
${\mathcal{M}} _{\pm}$ the corresponding signs $\epsilon _{\pm}$ determine
if the normal to the brane points in the direction of increasing ($\epsilon = +1$)
or decreasing ($\epsilon = -1$) radius.
\begin{figure}
\begin{center}
\includegraphics[width=8.5cm]{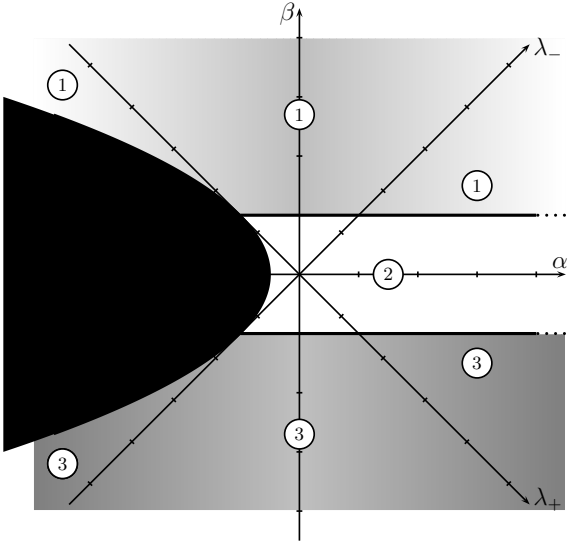}
\end{center}
\caption{\label{fig:epssig}The parameter space can be subdivided in three other regions
(strips) according to the values of the signs $\epsilon _{\pm}$. In region $1$ we have
$\epsilon _{\pm} = -1$, in region $2$ we have $\epsilon _{\pm} = \mp 1$ and, finally,
in region $3$ we have $\epsilon _{\pm} = +1$.}
\end{figure}
Using the results in (\ref{eq:claequequ}) for $\epsilon _{\pm}$ we can
subdivide the parameter space in three main regions, as in figure \ref{fig:epssig}.
These regions correspond to the following situations:
\begin{description}
    \item[region 1)] $\epsilon _{-} = \epsilon_{+} = -1$, so that in both spacetimes
    ${\mathcal{M} _{\pm}}$ the normal $n _{a} | _{\pm}$ to the brane trajectory points in the direction of
    decreasing $r$;
    \item[region 2)] $\epsilon_{-} = +1$ but $\epsilon_{+} = -1$, and in ${\mathcal{M}} _{-}$
    the normal to the brane trajectory $n _{a} | _{-}$ points in the direction of increasing
    $r _{-}$ but in ${\mathcal{M}} _{+}$ the normal $n _{a} | _{+}$ to the brane trajectory
    points in the direction of decreasing $r _{+}$;
    \item[region 3)] $\epsilon _{-} = \epsilon _{+} = +1$; thus in both spacetimes
    ${\mathcal{M} _{\pm}}$ the normals $n _{a} | _{\pm}$ to the brane trajectory
    point in the direction of increasing $r _{\pm}$.
\end{description}
We thus have various combinations of the geometries of the two spacetimes ${\mathcal{M} _{\pm}}$
according to the classification in figure \ref{fig:parspa}; moreover we have to combine them
choosing the part of spacetime on the correct side of the brane trajectory following the
classification in figure \ref{fig:epssig}. This gives a total of ten subcases, which are
summarized in figure \ref{fig:difcas}.
\begin{figure}
\begin{center}
\includegraphics[width=8.5cm]{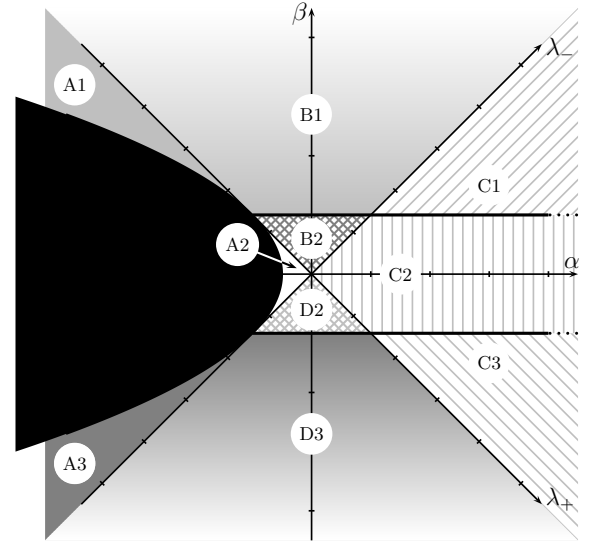}
\end{center}
\caption{\label{fig:difcas}In principle we have to study 10 different cases, corresponding to
as many different regions in the parameter space. These regions are obtained combining
the classifications in figures \ref{fig:parspa} and \ref{fig:epssig}, so that in each
region there is a well defined type of junction with an unique choice for the signs. As
we will see each of the pairs A1 and A3, B1 and D3, B2 and D2, C1 and C3 will correspond to
a distinct processes in spacetime; this matches well with the analytical result, in which the
action turns out to be an even function of $\beta$, and gives only a total of 6
different possible processes.}
\end{figure}
The naming convention is the most natural, so that, for instance, the junction named
B2 is a junction of type B, \textit{i.e.\/} a $\mathrm{dS} _{(-)}-\mathrm{AdS} _{(+)}$ junction,
with the signs as in region 2, \textit{i.e.\/} $\epsilon_{-} = +1$ and $\epsilon_{+} = -1$. All
other names follow the same convention and we thus obtain solutions of the following types
A1, A2, A3, B1, B2, C1, C2, C3, D2, D3; their causal structure is detailed in the following appendix.

\section{\label{app:clasol}Penrose diagrams \& Tunnelling}

Here, following the classification given in the previous appendix,
we show the corresponding Penrose diagrams for the
classical solutions, and a pictorial representation of the
corresponding tunnelling processes. For each case we give four
diagrams:
\begin{enumerate}
\item
the complete spacetime from which we have to ``cut'' the ${\mathcal{M}} _{-}$
part of the bulk (with the trajectory
of the bubble and the associated normal) in the top left diagram;
\item the complete spacetime from which we have to
``cut'' the ${\mathcal{M}} _{+}$ part of the bulk (with the trajectory
of the bubble and the associated normal) in the top right diagram;
\item the junction, \textit{i.e.\/} the full manifold ${\mathcal{M}}$ (again with
the shell trajectory and the corresponding normal) in the bottom left diagram;
\item a pictorial representation of the creation of the brane \emph{via} a tunnelling process,
in the bottom right diagram; in this case we have used time translation invariance to
set the ``tunnelling time'' at the coordinate time $t = 0$ so that the top half of the
diagram represents the final state (i.e. the top half of the junction in the bottom left
diagram) whereas the bottom half of the diagram is a representation (detailed below)
of the pre-tunnelling configuration.
\end{enumerate}

We would like to discuss preliminarily in more detail the way in which the pre-tunnelling configuration
is constructed. We remember that we are considering non negligible quantum effects
in the regime of the dynamics before the tunnelling. We will thus use for the signs
the results coming from the modified junction (\ref{eq:modadijuncon}), which are
those in (\ref{eq:modsig}). This gives a spacetime structure consisting of two
regions (which can be of the de Sitter or of the anti--de Sitter type depending
on the values of $\alpha$ and $\beta$) both bounded by $r = 0$ and by the shell
radius $r = R (\tau)$: this is the junction corresponding to the bounded solution of the modified
potential (\ref{eq:modeffpot}), which is a brane starting from zero
radius and expanding to a maximum radius (much smaller than any other length scale
present in the problem) before recollapsing to $r=0$.
In the limit in which $\bar{a} \to 0$, as discussed in section \ref{sec:tunspatimstr},
this maximum radius tends, in fact, to zero. Nevertheless, in our representation
of the spacetime before the tunnelling we have kept an arbitrarily finite size
for the maximum radius, to make the diagram more readable. At the same time,
we have slightly ``blurred'' it to make pictorially
explicit that we are not dealing with a purely classical configuration but with a somehow
``heuristic'' representation of a spacetime where quantum effects are highly
non trivial and certainly non negligible. We would like, anyway, to stress again
that these quantum configurations will be the initial state of tunnelling processes that correspond
to what in the literature has also been called ``tunnelling from nothing'' (see
for instance \cite{ansoldi1} and references therein).

We will now present in detail the various kinds of junctions.

\subsection{Type A}

\begin{figure}[!ht]
\begin{center}
\begin{tabular}{|c|c|}
\hline
\includegraphics[height=4cm]{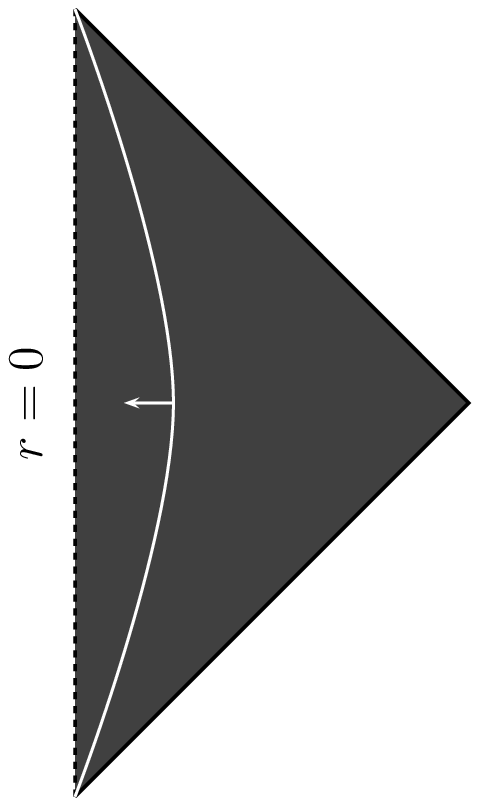}
&
\includegraphics[height=4cm]{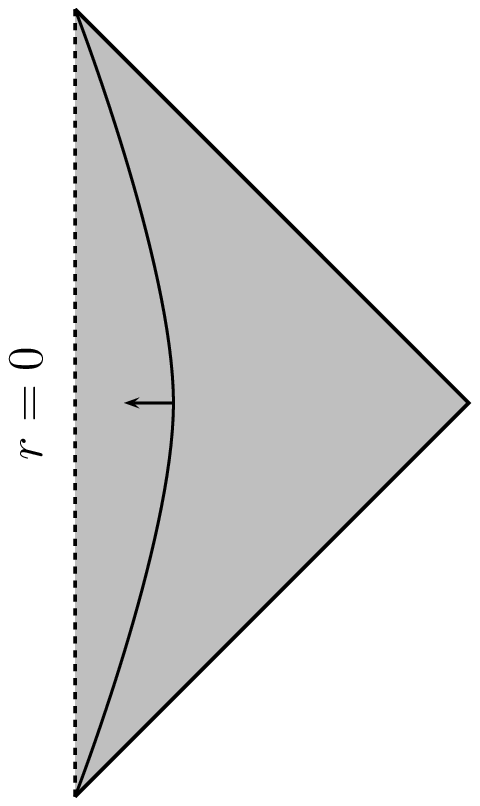}
\cr $\mathrm{AdS} _{-} , \epsilon _{-}=-1$ & $\mathrm{AdS} _{+} , \epsilon _{+}=-1$ \cr
\hline
\includegraphics[height=4cm]{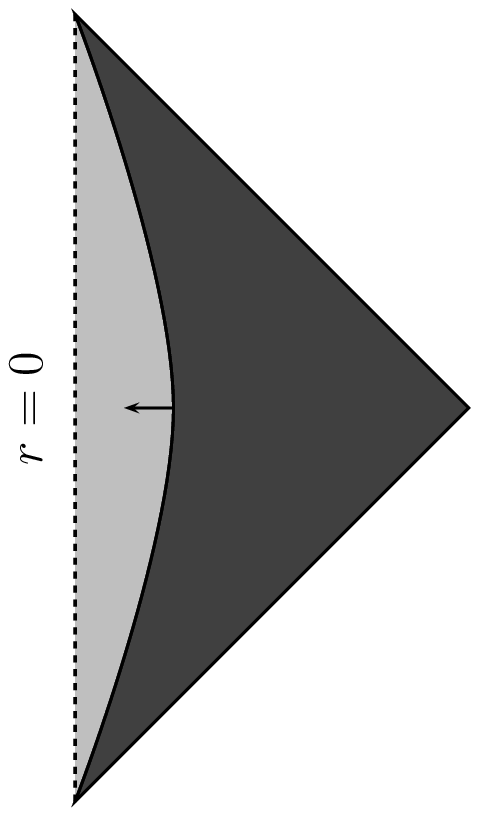}
&
\includegraphics[height=4cm]{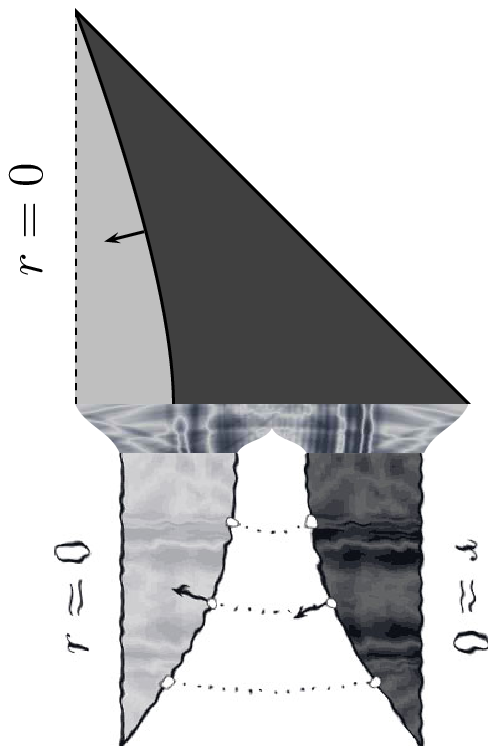}
\cr
$\mathrm{AdS} _{+}/\mathrm{AdS} _{-}$ & Tunnelling
\cr
\hline
\end{tabular}
\caption{\label{fig:junA_1}Case A1: AdS/AdS, $\epsilon_{-}=\epsilon_{+}=-1$}
\end{center}
\end{figure}

\begin{figure}[!ht]
\begin{center}
\begin{tabular}{|c|c|}
\hline
\includegraphics[height=4cm]{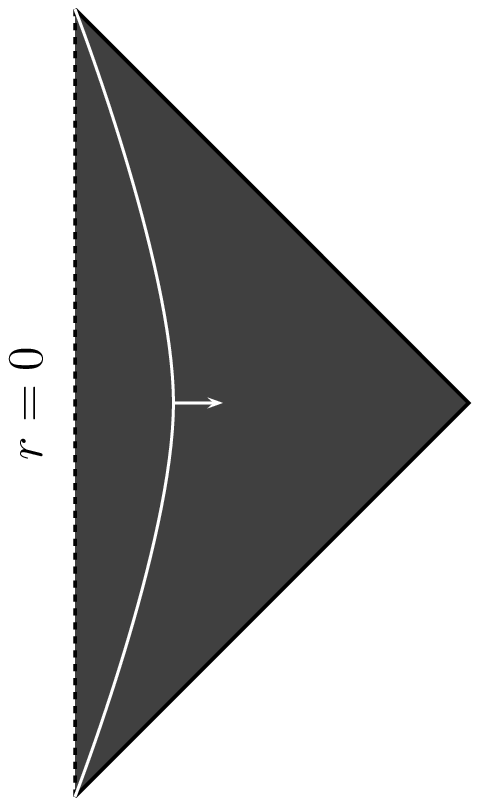}
&
\includegraphics[height=4cm]{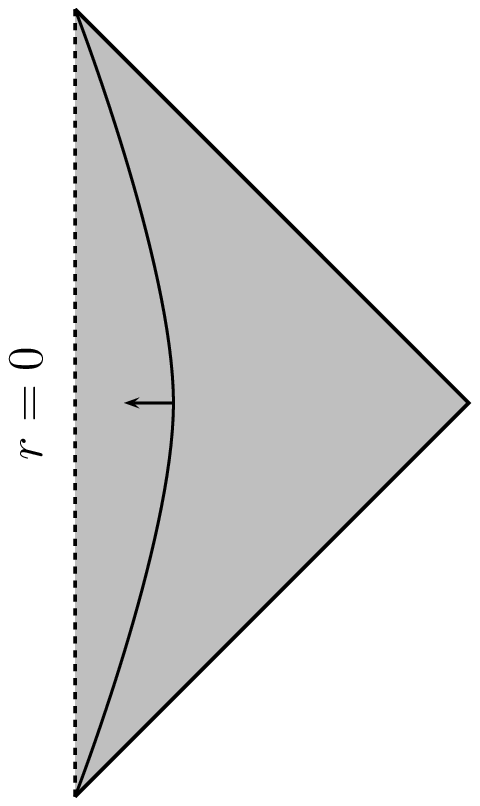}
\cr $\mathrm{AdS} _{-} , \epsilon _{-}=+1$ & $\mathrm{AdS} _{+} , \epsilon _{+}=-1$ \cr
\hline
\includegraphics[height=4cm]{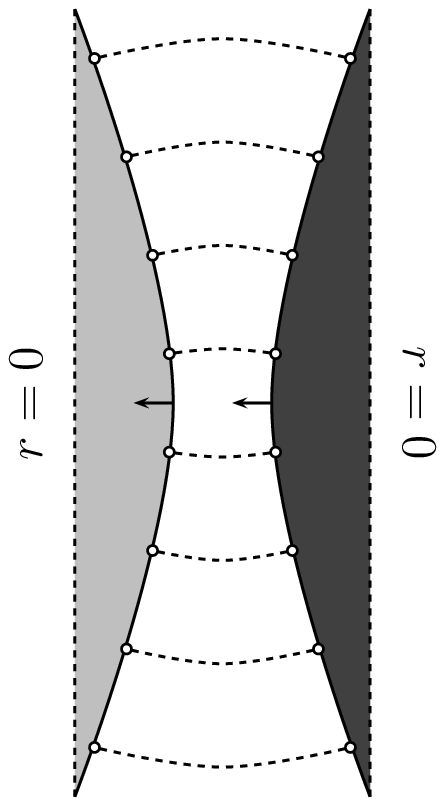}
&
\includegraphics[height=4cm]{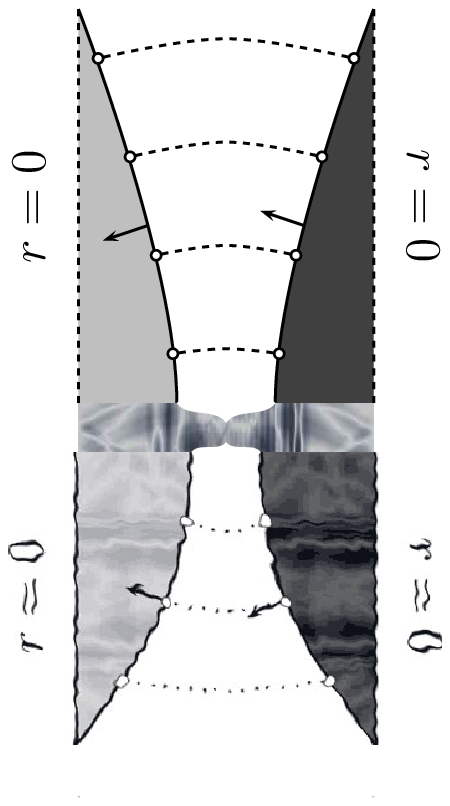}
\cr
$\mathrm{AdS} _{+}/\mathrm{AdS} _{-}$ & Tunnelling
\cr
\hline
\end{tabular}
\caption{\label{fig:junA_2}Case A2: AdS/AdS, $\epsilon_{-}=+1;\epsilon_{+}=-1$}
\end{center}
\end{figure}

\begin{figure}[!ht]
\begin{center}
\begin{tabular}{|c|c|}
\hline
\includegraphics[height=4cm]{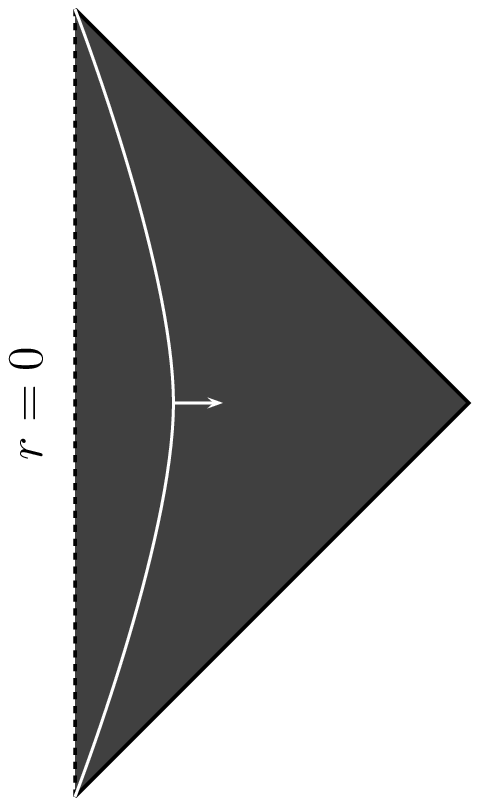}
&
\includegraphics[height=4cm]{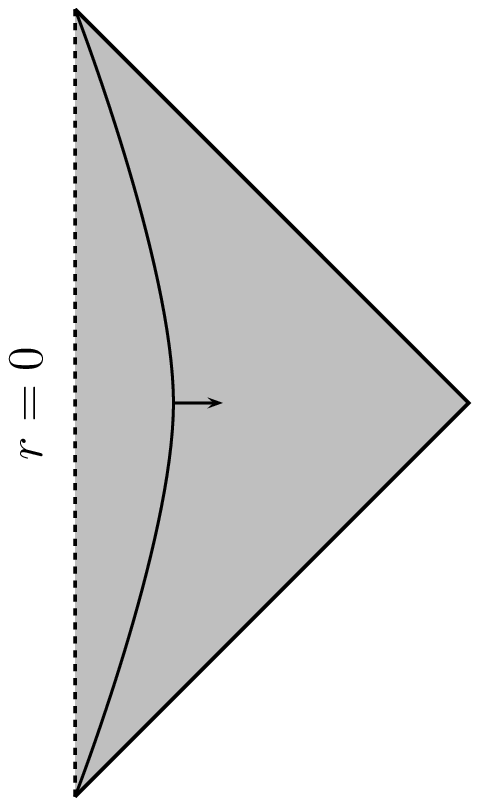}
\cr $\mathrm{AdS} _{-} , \epsilon _{-}=+1$ & $\mathrm{AdS} _{+} , \epsilon _{+}=+1$ \cr
\hline
\includegraphics[height=4cm]{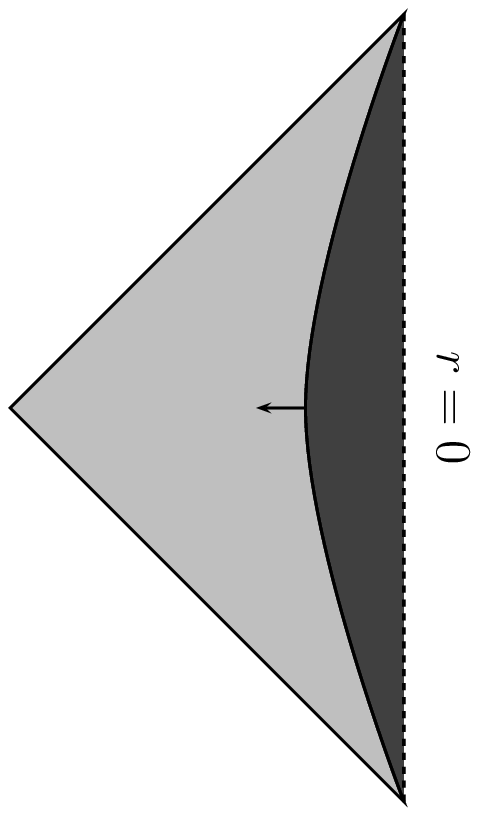}
&
\includegraphics[height=4cm]{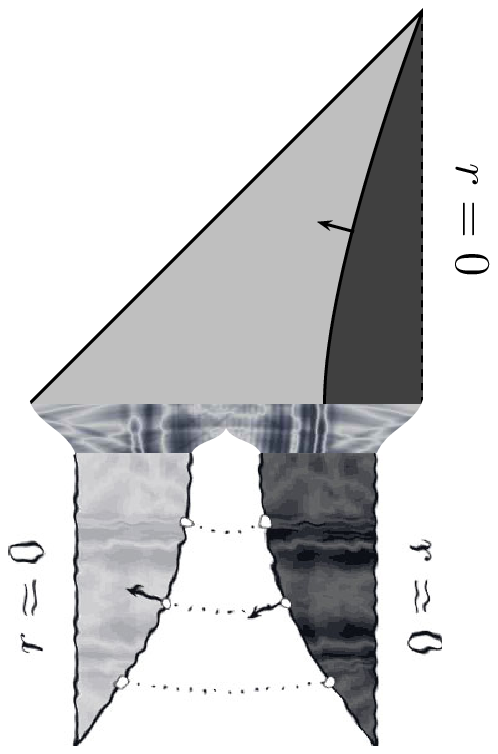}
\cr
$\mathrm{AdS} _{+}/\mathrm{AdS} _{-}$ & Tunnelling
\cr
\hline
\end{tabular}
\caption{\label{fig:junA_3}Case A3: AdS/AdS, $\epsilon_{-}=\epsilon_{+}=+1$}
\end{center}
\end{figure}

As discussed in main text (and with reference to figure \ref{fig:difcas}),
this class of junctions consists of the matching of two
anti--de Sitter spacetimes with different cosmological constants. There are three
possibilities corresponding to type A1, A2 and A3 junctions, which are
shown, respectively, in figures \ref{fig:junA_1}, \ref{fig:junA_2} and \ref{fig:junA_3}.

The junction of type A1 (figure \ref{fig:junA_1}) has $\epsilon _{\pm} = -1$,
so in both anti-de Sitter spacetimes the normal to the brane points in the direction of decreasing
radii. For this junction (and for all the ones in the 1 sector),
$\beta > 1$, so that $\lambda _{-}$ is always greater than
$\lambda _{+}$. The classical junction is seen (by an observer travelling toward increasing
values of the radius) as a transition from a more negative to a less negative value of the
cosmological constant, when he/she crosses the brane. The global picture of the tunnelling
process is then obtained according to the prescription discussed in section \ref{sec:tunspatimstr}.
The pre-tunnelling spacetime has been discussed above and now consists of two ``small''
parts of anti--de Sitter spacetime with different cosmological constant. The tunnelling process
shows the transition between a compact spacetime composed by two anti-de Sitter regions
to a spacetime similar to anti--de Sitter, except for the fact that at some radius
the cosmological constant changes its value.

The junction of type A2, is instead a junction where effectively a tiny junction of
two anti-de Sitter spacetimes is ``inflated'' (figure \ref{fig:junA_2}).
In this case the signs are given by
$\epsilon _{\pm} = \mp 1$. Thus the normal to the brane points
in the direction of increasing radii in the anti--de Sitter spacetime with
cosmological constant $\lambda _{-}$, but it points in the direction of
decreasing radii in the anti--de Sitter spacetime with cosmological constant
$\lambda _{+}$. The net effect of the tunnelling process, which starts from the
already discussed initial state, is thus to inflate the pre-tunnelling compact
spacetime into a similar, but much larger, one (note that for graphical reasons the
scales in the pre-tunnelling and post-tunnelling parts of the diagram are different;
we remember that the maximum radius of the pre-tunnelling state will be much smaller
than any other length scale in the problem). The various diagrams are in
figure \ref{fig:junA_2}.

The final type A diagram is the A3 one. This is very much as the A1 type,
only that now $\epsilon _{\pm} = +1$; in both spacetime
the normal to the shell trajectory points in the direction of increasing
radii. Thus (apparently)
the role of ${\mathcal{M}} _{\pm}$ is interchanged,
as shown in figure \ref{fig:junA_3} where the dark and light gray parts
looks complementary to those in figure \ref{fig:junA_1}; on the other
hand, now we have $\beta < -1$ so that
$\lambda _{+} > \lambda _{-}$. The global spacetime structure
in the bottom left corner of figure \ref{fig:junA_3} implies that
an observer crossing the brane in the direction of increasing
values of the radius, perceives again a transition from a more negative
to a less negative value of the cosmological constant. So this diagram
describes exactly the same process described by type A1, as it is possible to
see from the diagram in figure \ref{fig:junA_3}.

\subsection{Type B}

The type B solutions, B1 and B2, correspond to the junction of a part
of a de Sitter spacetime, ${\mathcal{M}} _{-}$, with a part of
anti--de Sitter spacetime, ${\mathcal{M}} _{+}$. Although the pre-tunnelling
picture does not change very much, the configurations before the tunnelling
are now de Sitter--anti-de Sitter junctions.

\begin{figure}[!ht]
\begin{center}
\begin{tabular}{|c|c|}
\hline
\includegraphics[height=4cm]{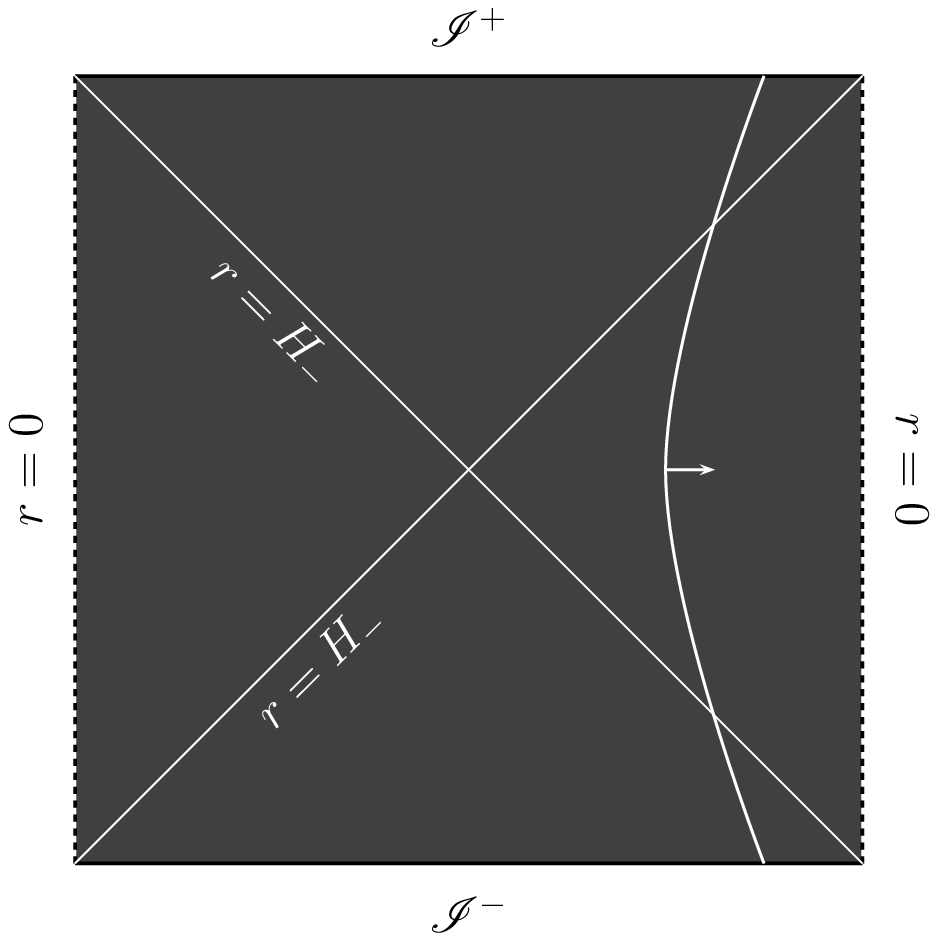}
&
\includegraphics[height=4cm]{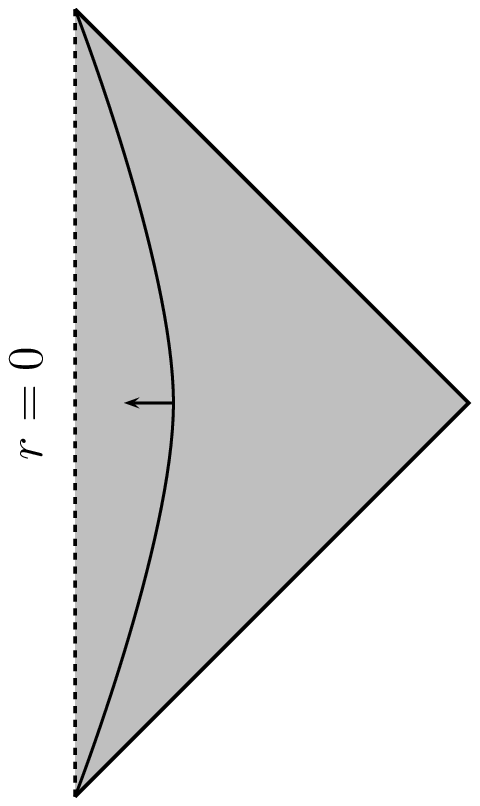}
\cr $\mathrm{dS} _{-} , \epsilon _{-}=-1$ & $\mathrm{AdS} _{+} , \epsilon _{+}=-1$ \cr
\hline
\includegraphics[height=4cm]{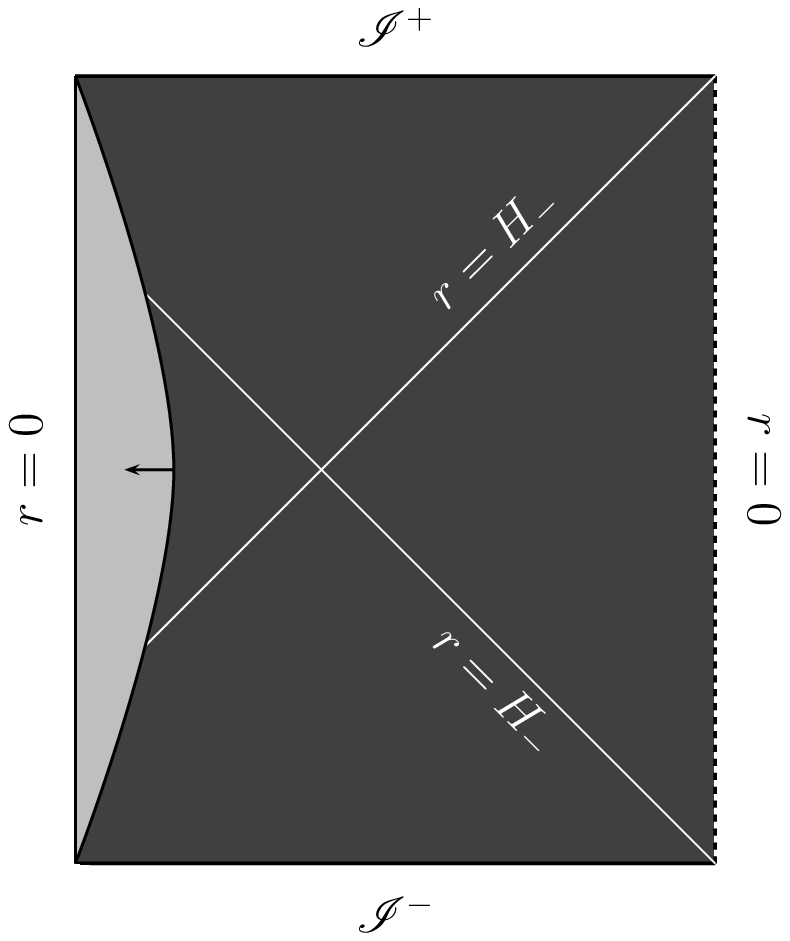}
&
\includegraphics[height=4cm]{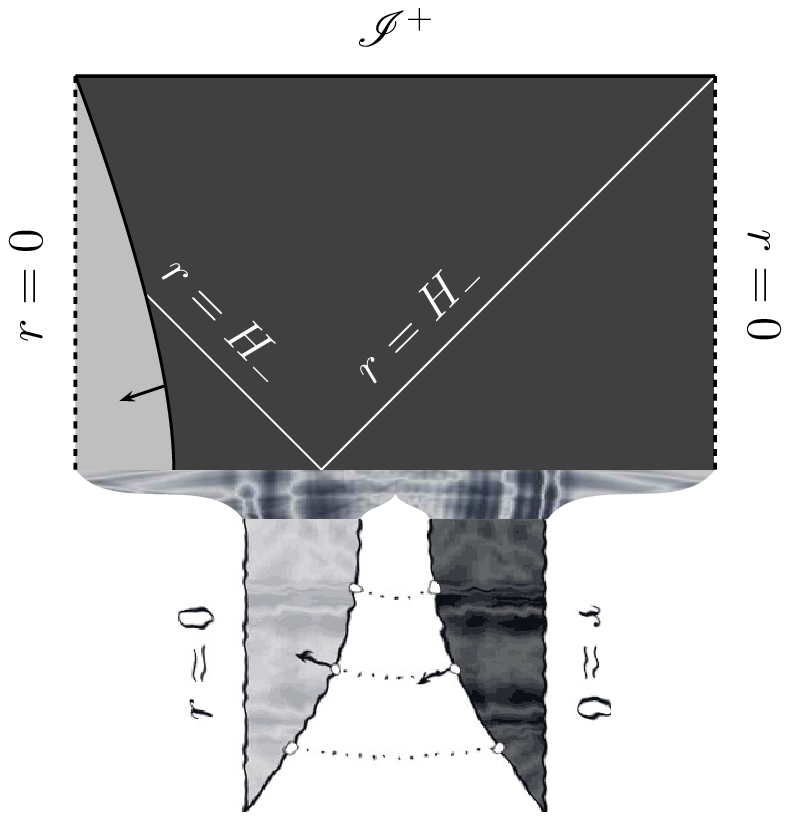}
\cr
$\mathrm{AdS} _{+}/\mathrm{dS} _{-}$ & Tunnelling
\cr
\hline
\end{tabular}
\caption{\label{fig:junB_1}Case B1: dS/AdS, $\epsilon_{-}=\epsilon_{+}=-1$}
\end{center}
\end{figure}

Let us first discuss the case B1, shown in figure \ref{fig:junB_1}. In both spacetimes,
the normal to the shell points in the direction of decreasing radii, since we
have $\epsilon _{\pm} = -1$. The ${\mathcal{M}} _{-}$ part of spacetime corresponds
to the region of de Sitter complementary to the region between $r=0$ and the brane's world-volume,
whereas the anti--de Sitter
part ${\mathcal{M}} _{+}$ is the bounded region in the upper right picture of
figure \ref{fig:junB_1}. Thus the spacetime ${\mathcal{M}}$ shows a transition from
a negative to a positive value of the cosmological constant for an observer
crossing the shell in the direction of increasing radius. The configuration before the tunnelling
has been described above; the tunnelling process again ``inflates'' an anti--de Sitter---de Sitter
junction with small volume to a much larger one, as shown in the bottom right diagram of
figure \ref{fig:junB_1}.

\begin{figure}[!ht]
\begin{center}
\begin{tabular}{|c|c|}
\hline
\includegraphics[height=4cm]{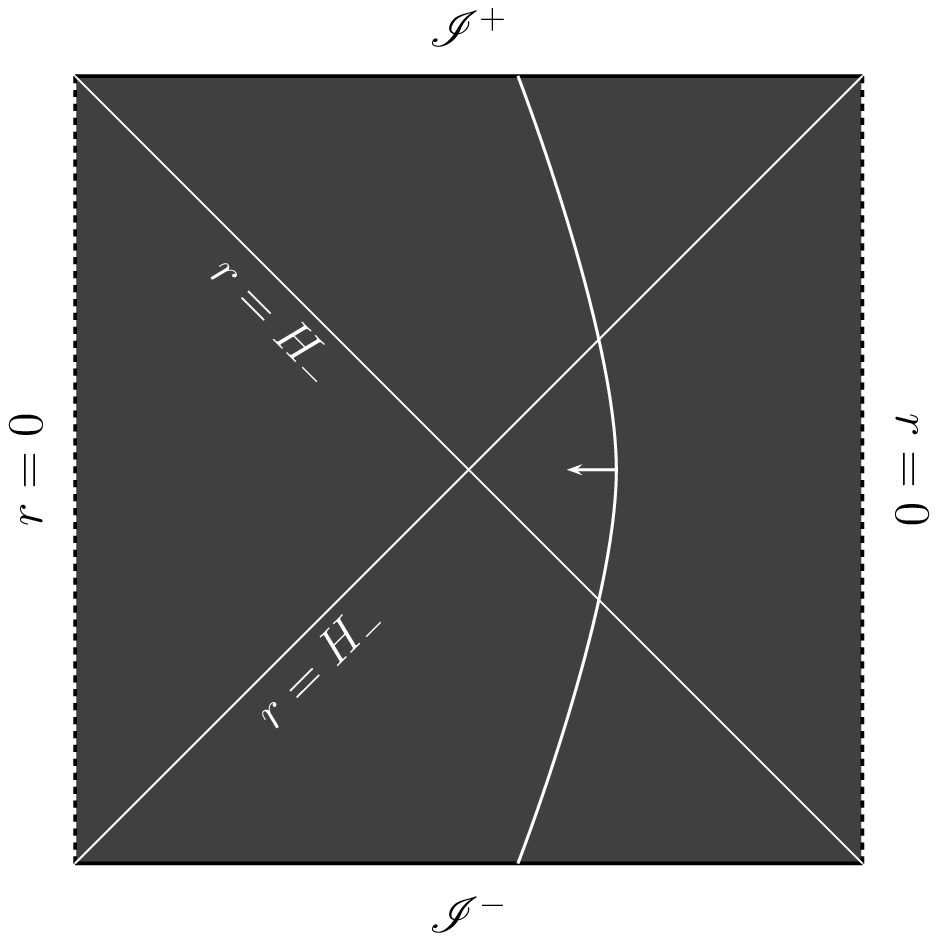}
&
\includegraphics[height=4cm]{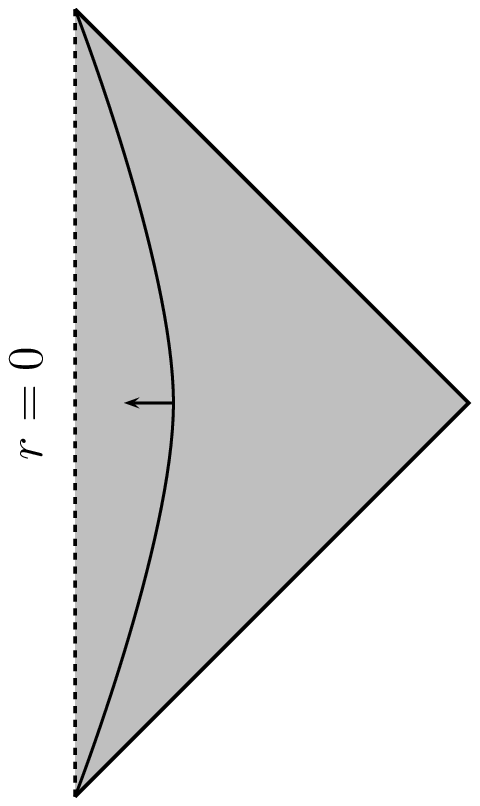}
\cr $\mathrm{dS} _{-} , \epsilon _{-}=+1$ & $\mathrm{AdS} _{+} , \epsilon _{+}=-1$ \cr
\hline
\includegraphics[height=4cm]{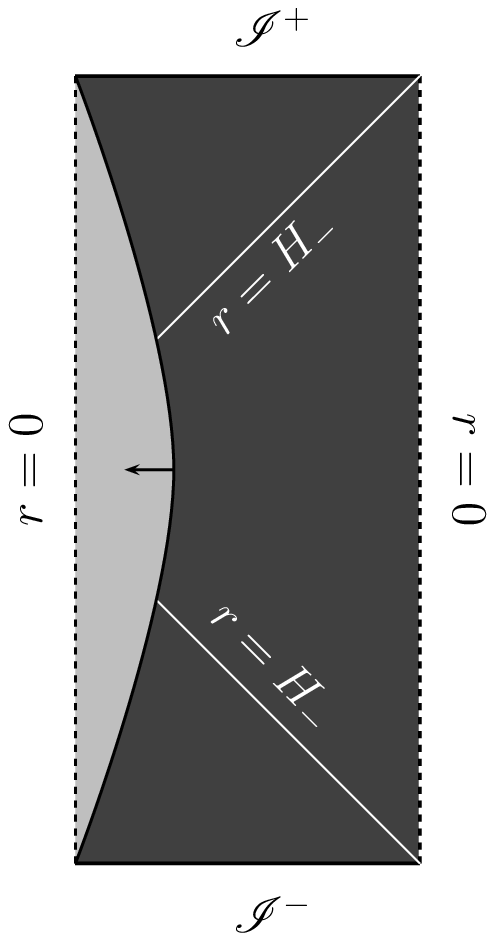}
&
\includegraphics[height=4cm]{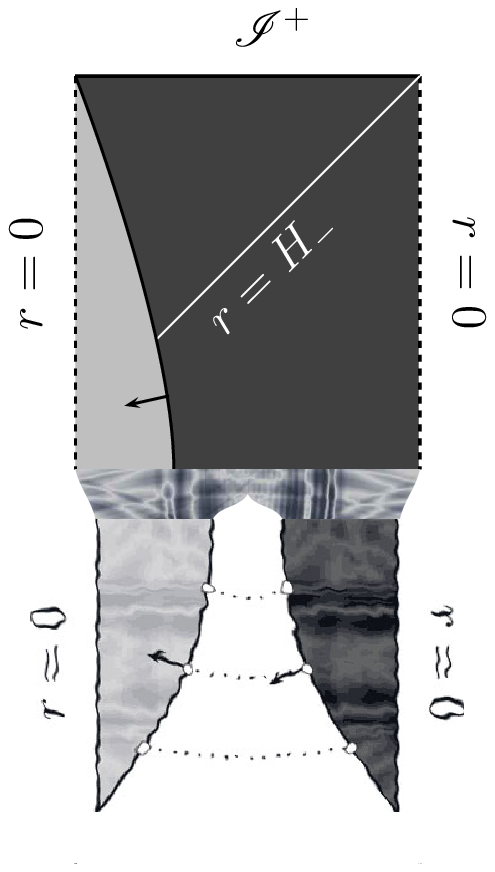}
\cr
$\mathrm{AdS} _{+}/\mathrm{dS} _{-}$ & Tunnelling
\cr
\hline
\end{tabular}
\caption{\label{fig:junB_2}Case B2: dS/AdS, $\epsilon_{-}=+1;\epsilon_{+}=-1$}
\end{center}
\end{figure}

We then come to the case B2, shown in figure \ref{fig:junB_2}. Since in this case
the signs which fix the orientation of the normals are $\epsilon _{\pm} = \mp 1$,
the situation for ${\mathcal{M}} _{+}$ is as in the previous case, but the
${\mathcal{M}} _{-}$ part of spacetime changes because of the change in the sign
of $\epsilon _{+}$. This gives for the junction the diagram in the bottom left part
of figure \ref{fig:junB_2}. The configuration before and after the tunnelling is shown
in the bottom right corner of figure \ref{fig:junB_2}. The situation is very similar
to the one in the previous case.

\subsection{Type C}

Type C solutions are the ``de Sitter counterpart'' of type A solution. The
main difference (not related to the spacetime structure) is that whereas for
anti--de Sitter junctions the values of the cosmological constant are restricted
(\textit{i.e.\/} given two arbitrary values a junction might not exists) de Sitter junctions
exists for all values of the cosmological constants $\lambda _{\pm}$ (this is
clear from figure \ref{fig:difcas}). Please, also remember that now the pre-tunnelling
diagram will consist of a junction of two de Sitter spacetime with different cosmological constants
and with all the other properties described above.

\begin{figure}[!ht]
\begin{center}
\begin{tabular}{|c|c|}
\hline
\includegraphics[height=4cm]{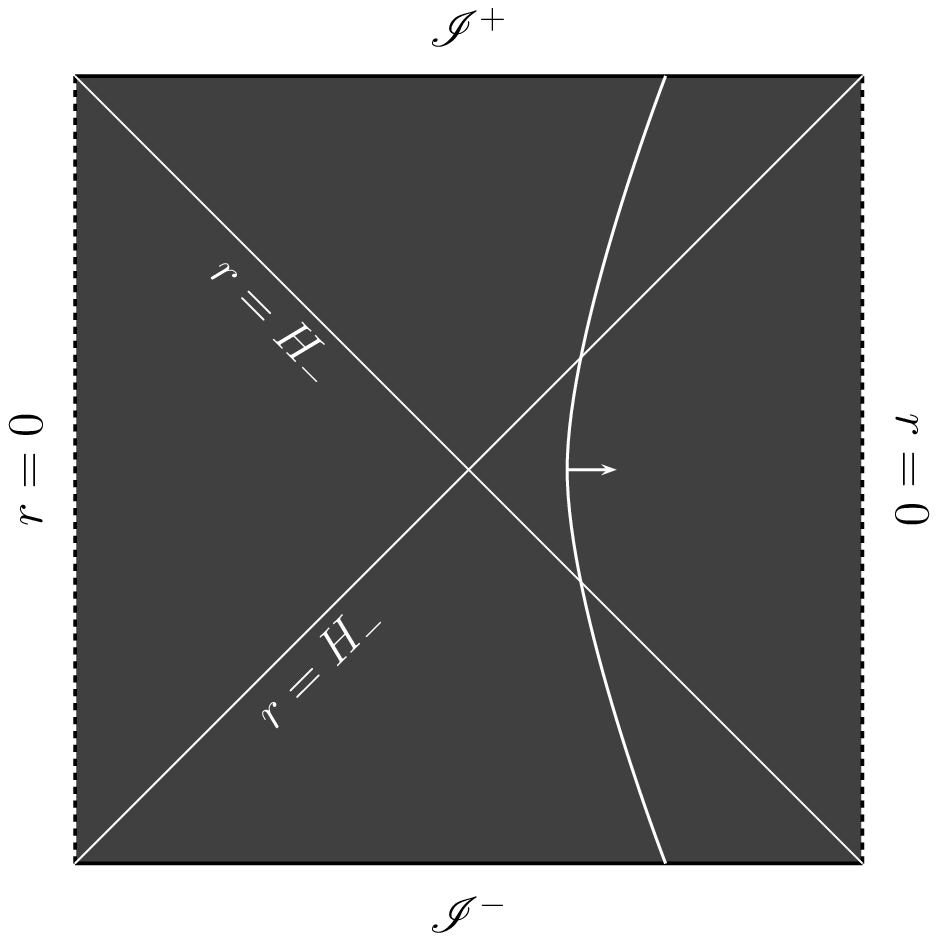}
&
\includegraphics[height=4cm]{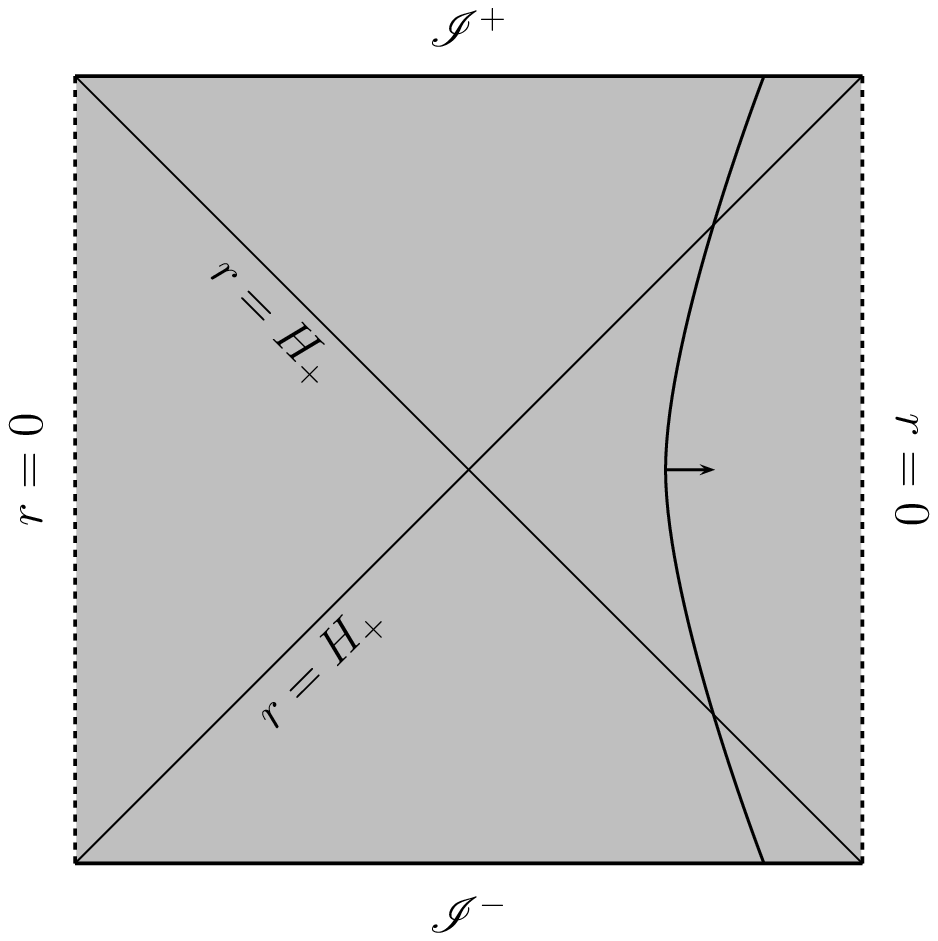}
\cr $\mathrm{dS} _{-} , \epsilon _{-}=-1$ & $\mathrm{dS} _{+} , \epsilon _{+}=-1$ \cr
\hline
\includegraphics[height=4cm]{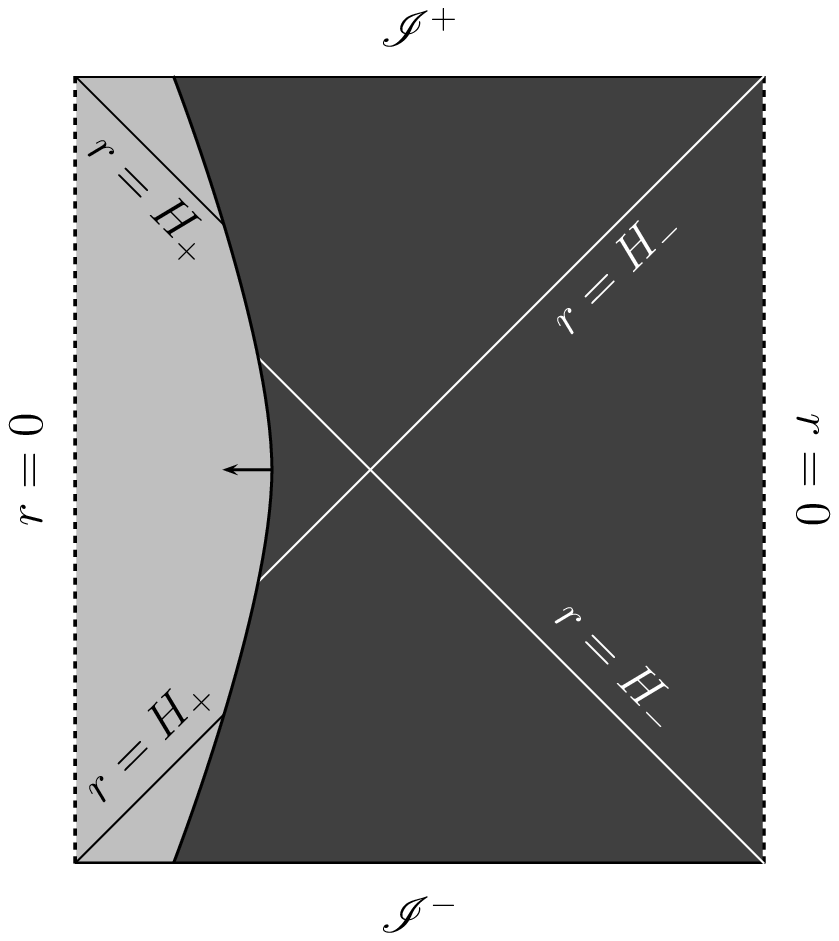}
&
\includegraphics[height=4cm]{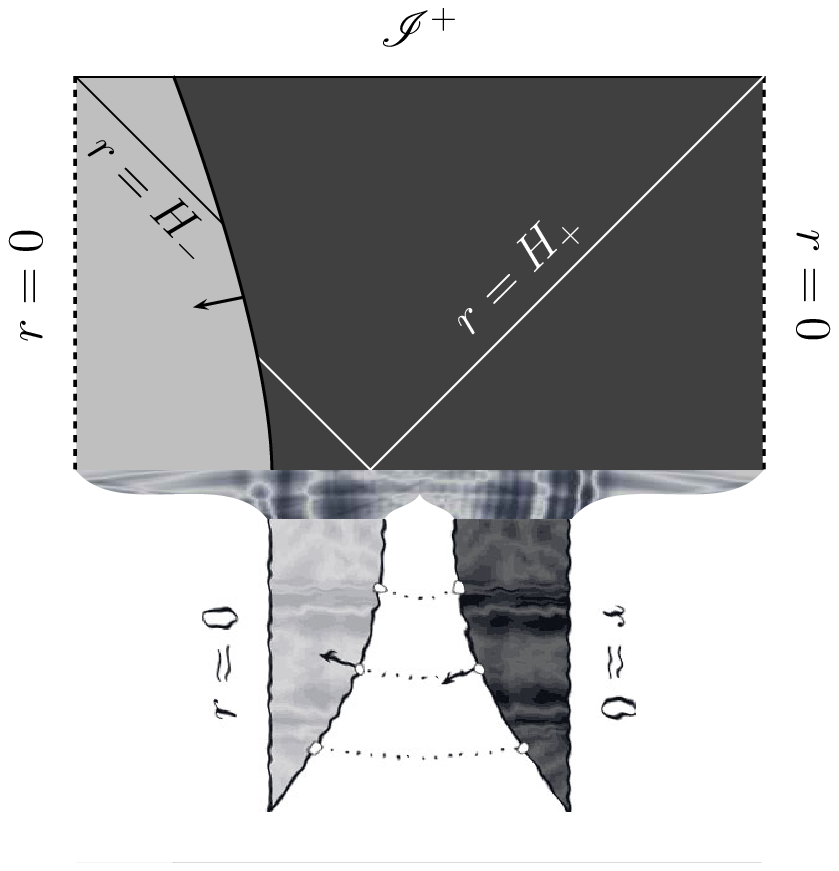}
\cr
$\mathrm{dS} _{+}/\mathrm{dS} _{-}$ & Tunnelling
\cr
\hline
\end{tabular}
\caption{\label{fig:junC_1}Case C1: dS/dS, $\epsilon_{-}=\epsilon_{+}=-1$}
\end{center}
\end{figure}

Let us start with case C1, shown in figure \ref{fig:junC_1}. For this junction,
the signs are $\epsilon _{\pm} = -1$, so in both de Sitter spacetimes the normals
to the brane point in the direction of decreasing radii. Moreover,
since in this case $\beta > 1$, the cosmological constant in ${\mathcal{M}} _{-}$
will be bigger than the cosmological constant in ${\mathcal{M}} _{+}$, so that
the cosmological horizon $H _{-}$ will be smaller than $H _{+}$. The diagram for
the junction is shown in the bottom left part of figure \ref{fig:junC_1}. It is
interesting to see some effects of the brane on the spacetime structure. There
are observers that crossing the brane can move behind the cosmological horizon
of ${\mathcal{M}} _{-}$ without going through $H _{-}$. They can also come out,
by crossing the brane in the opposite direction. At the same time observers,
by moving accurately and with proper timing, might end up behind the cosmological
horizon of ${\mathcal{M}} _{+}$ without crossing any horizon but using the
shell as a ``gate''\footnote{These considerations might be modified if we consider
the influence of the observer on spacetime. To neglect this influence, as a first approximation,
is fairly common in the literature to obtain some hints
about the global spacetime structure.}. As usual no additional considerations are
required for the before tunnelling configuration. We note, instead, that the semiclassical
transition has again the effect of ``inflating'' a small quantum spacetime region,
as clearly shown in figure \ref{fig:junC_1}.

\begin{figure}[!ht]
\begin{center}
\begin{tabular}{|c|c|}
\hline
\includegraphics[height=4cm]{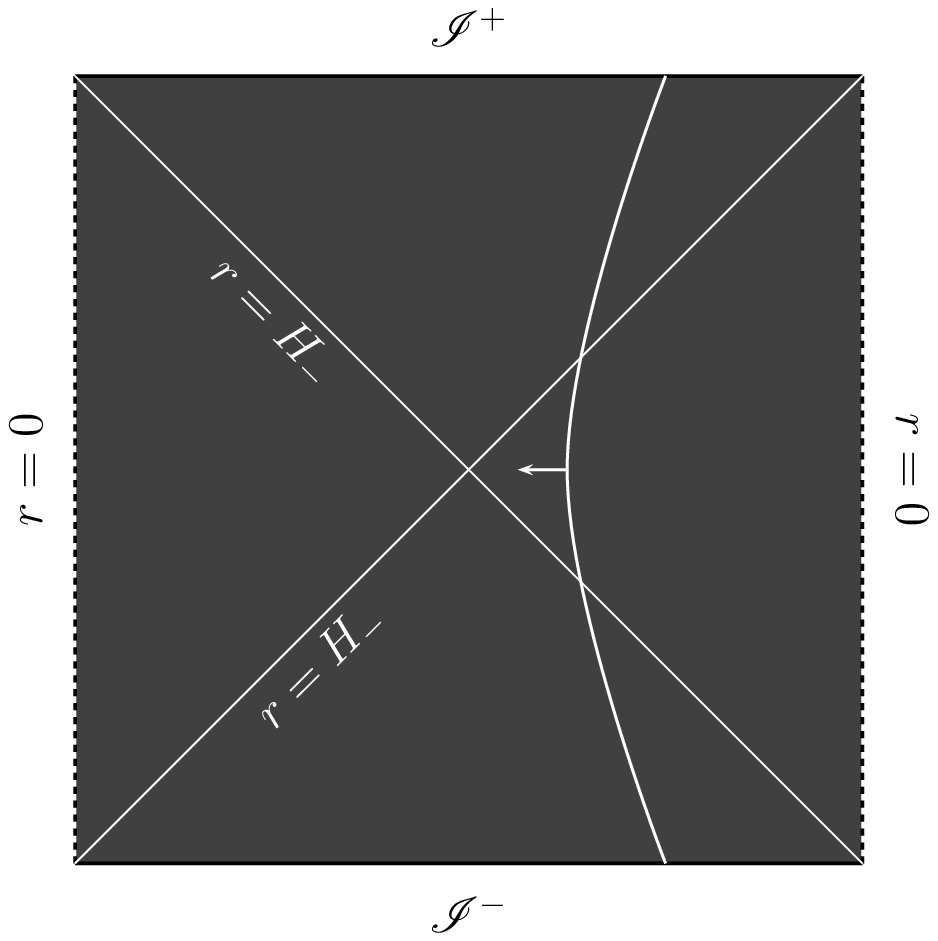}
&
\includegraphics[height=4cm]{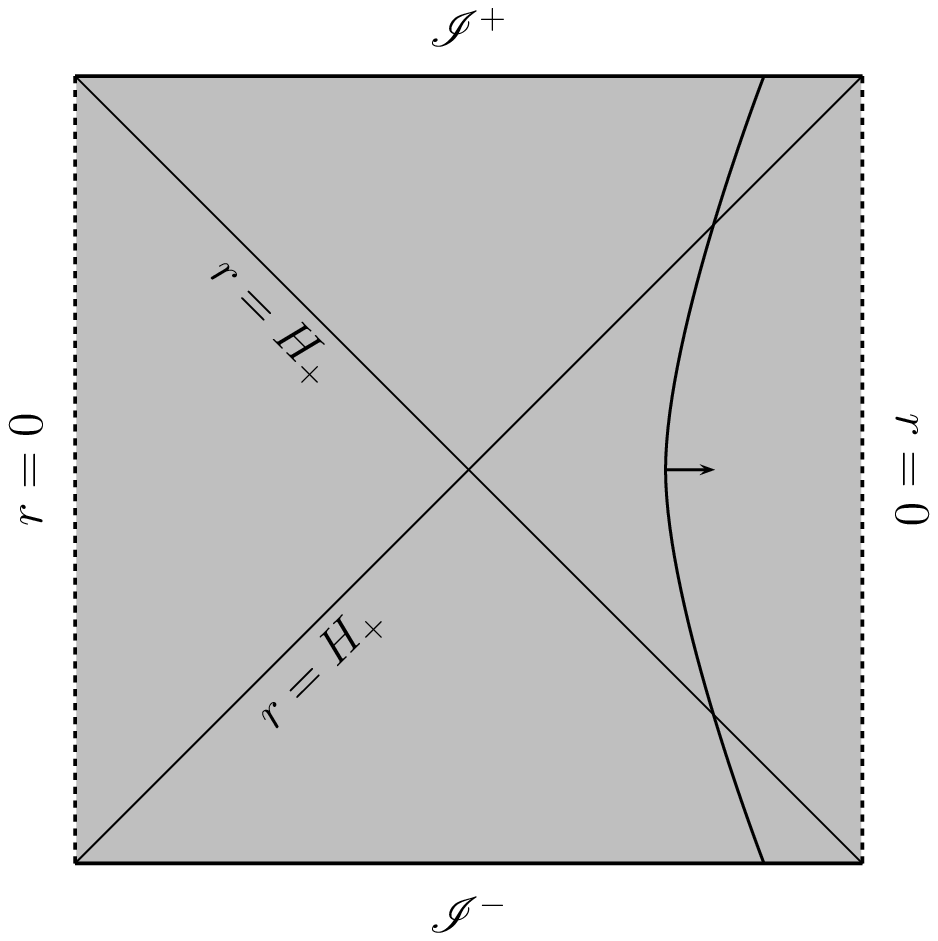}
\cr $\mathrm{dS} _{-} , \epsilon _{-}=+1$ & $\mathrm{dS} _{+} , \epsilon _{+}=-1$ \cr
\hline
\includegraphics[height=4cm]{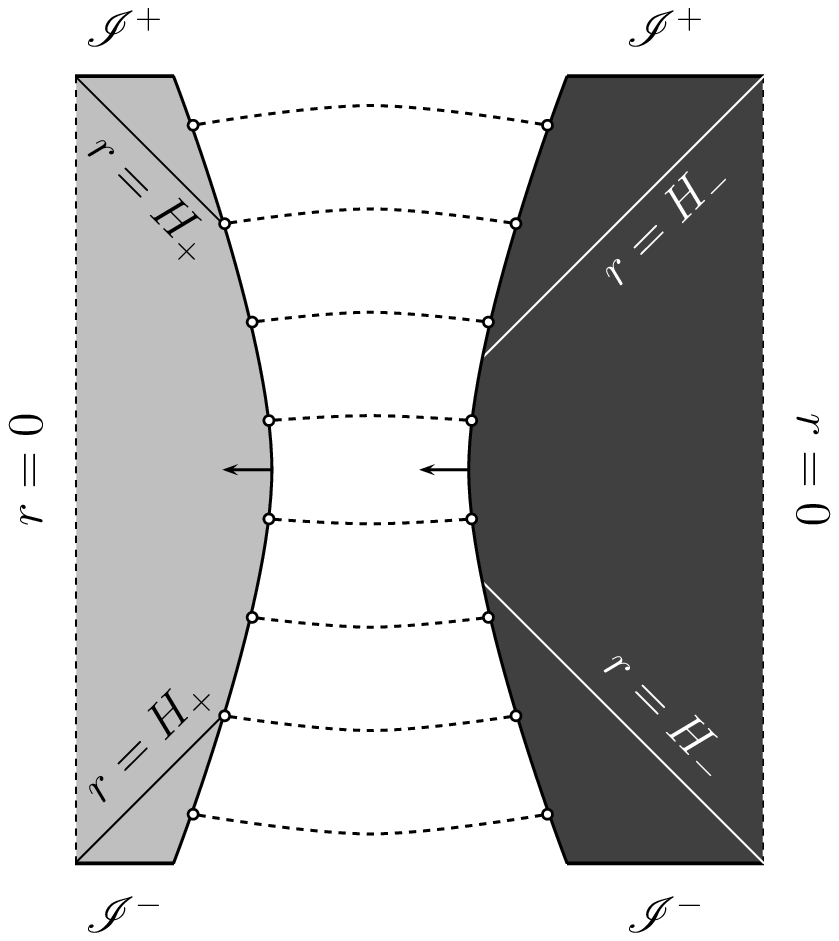}
&
\includegraphics[height=4cm]{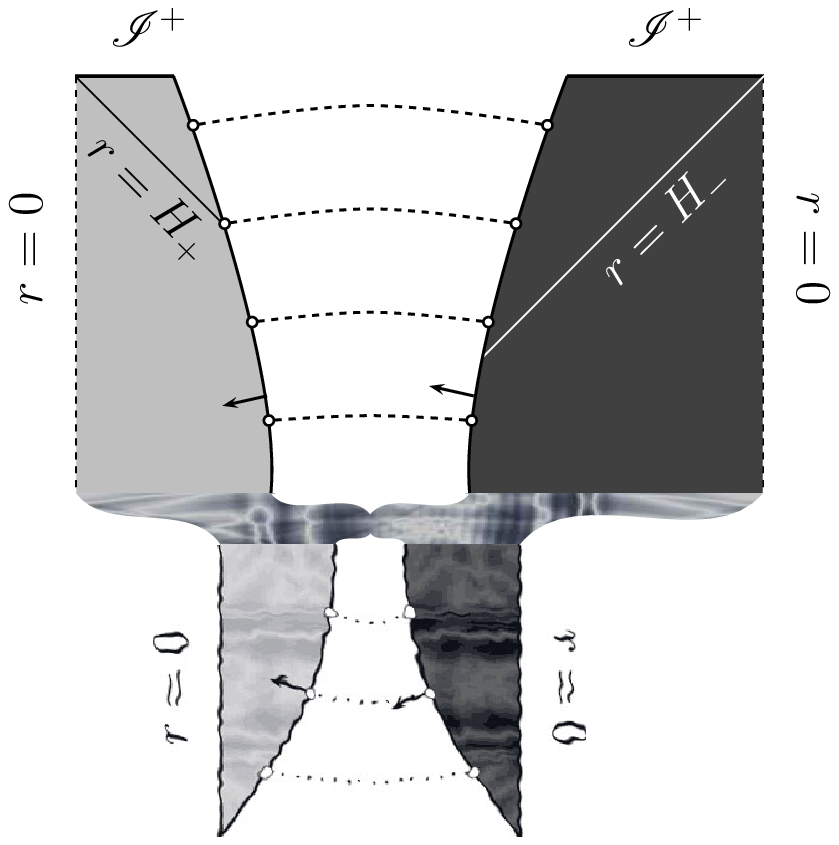}
\cr
$\mathrm{dS} _{+}/\mathrm{dS} _{-}$ & Tunnelling
\cr
\hline
\end{tabular}
\caption{\label{fig:junC_2}Case C2: dS/dS, $\epsilon_{-}=+1;\epsilon_{+}=-1$}
\end{center}
\end{figure}
We now discuss case C2, in which the signs are $\epsilon _{\pm} = \mp 1$. Then
the normal to the brane in the de Sitter space with cosmological constant
$\lambda _{-}$ points in the direction of increasing radii, whereas the normal
to the brane in the de Sitter space with cosmological constant $\lambda _{+}$
points in the direction of decreasing radii. We have again two regions of
spacetime of the de Sitter type but with a different cosmological constant
joined across the brane (bottom right part of figure \ref{fig:junC_2}).
The full picture of the tunnelling process requires the consideration
of the same initial configuration used in case C1 above.  We also point out
that in this case, the sign of $\beta$ is not fixed, so either of the cosmological
constants $\lambda _{\pm}$ may be the bigger. In particular, the junction in the bottom
left corner of figure \ref{fig:junC_2} shows the situation in which
$\lambda _{+} < \lambda _{-}$, so that $H _{+} > H _{-}$. In the
bottom right part of figure \ref{fig:junC_2} a representation of the tunnelling process is shown.

\begin{figure}[!ht]
\begin{center}
\begin{tabular}{|c|c|}
\hline
\includegraphics[height=4cm]{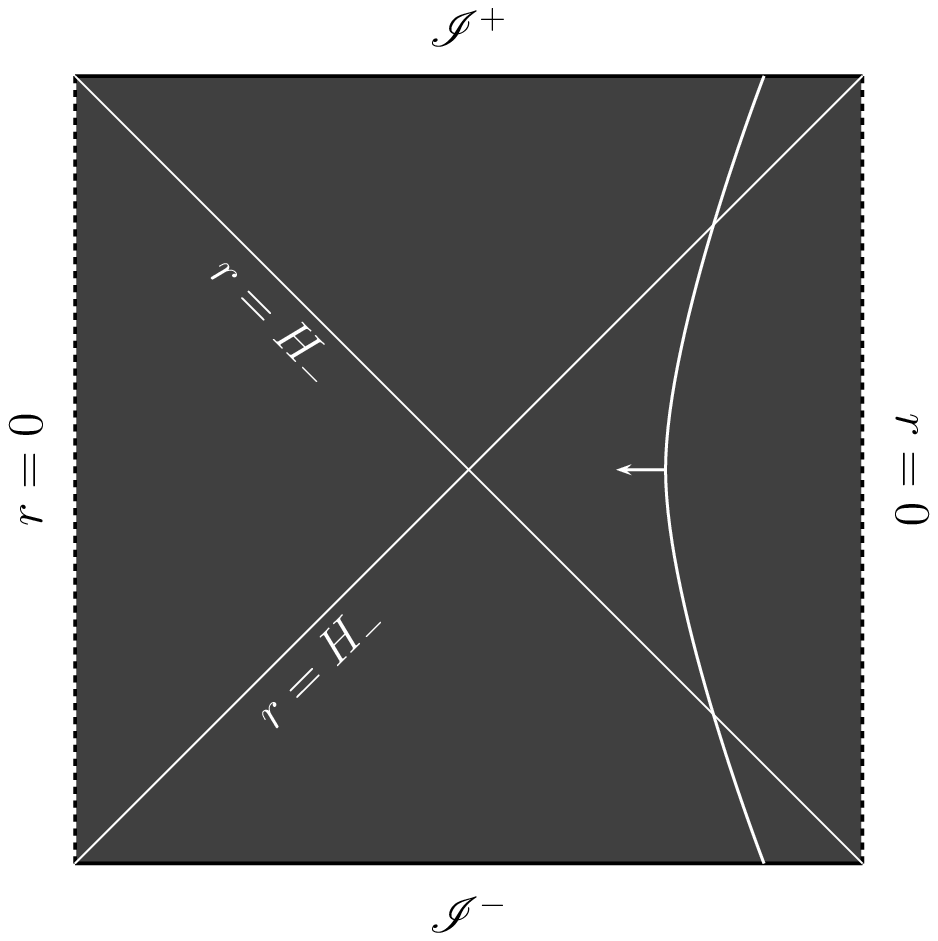}
&
\includegraphics[height=4cm]{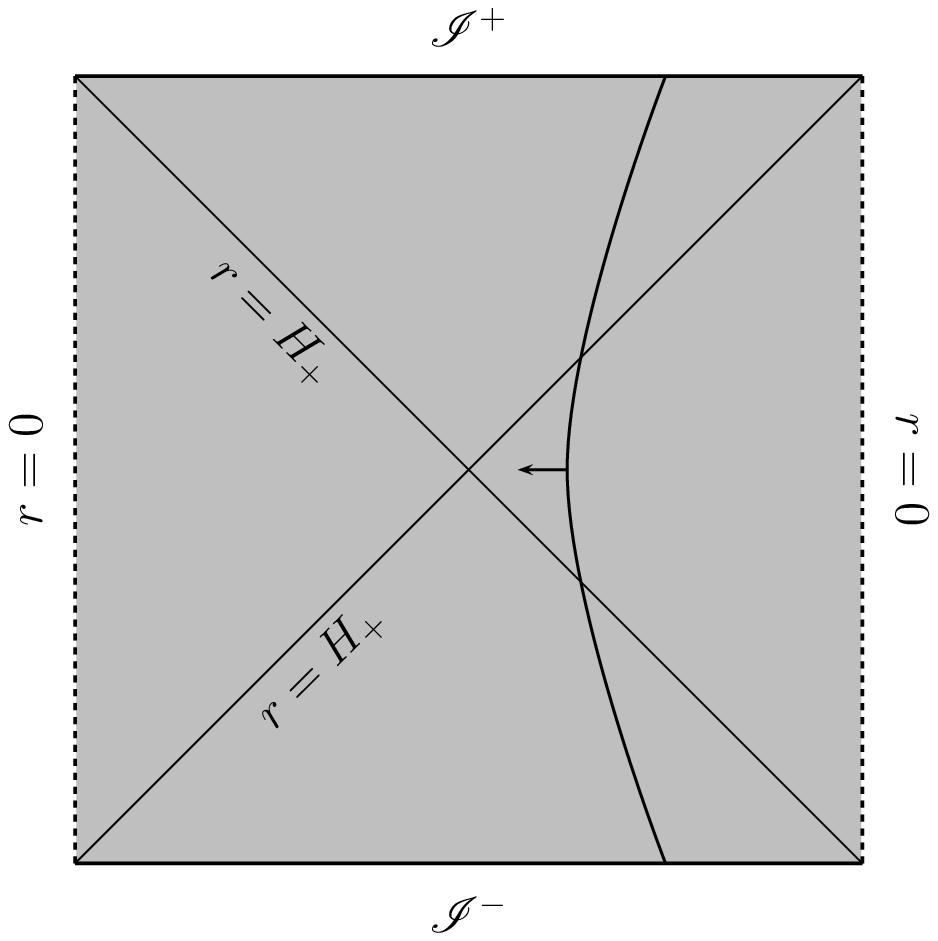}
\cr $\mathrm{dS} _{-} , \epsilon _{-}=+1$ & $\mathrm{dS} _{+} , \epsilon _{+}=+1$ \cr
\hline
\includegraphics[height=4cm]{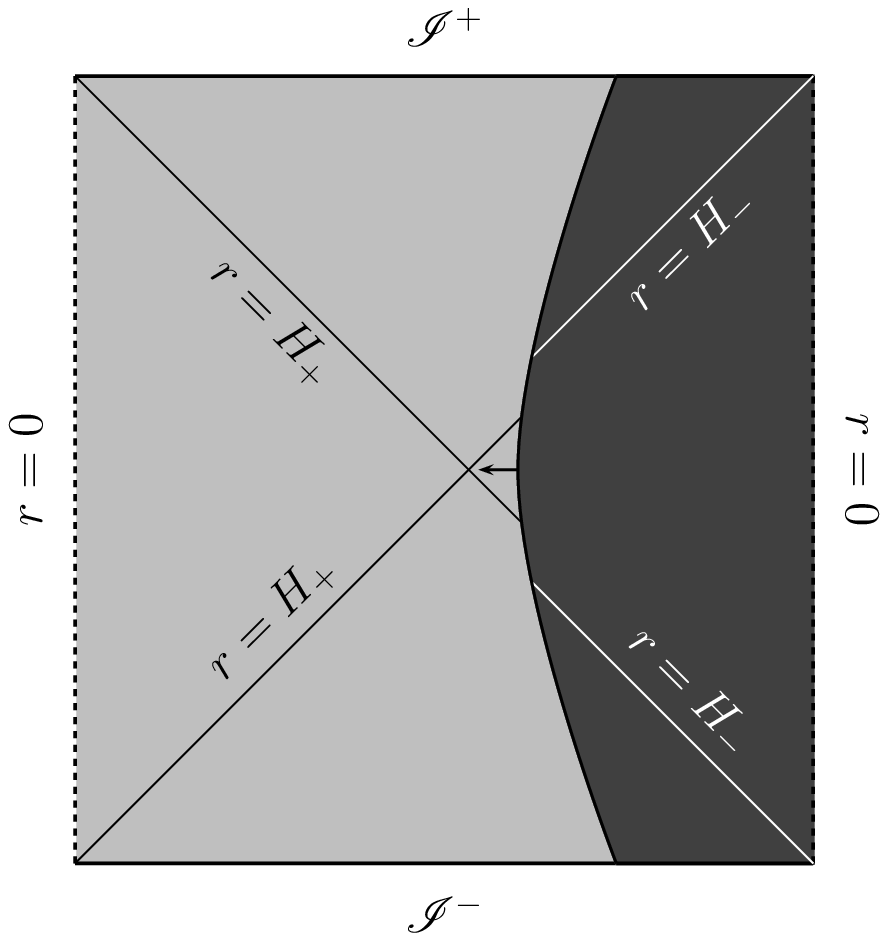}
&
\includegraphics[height=4cm]{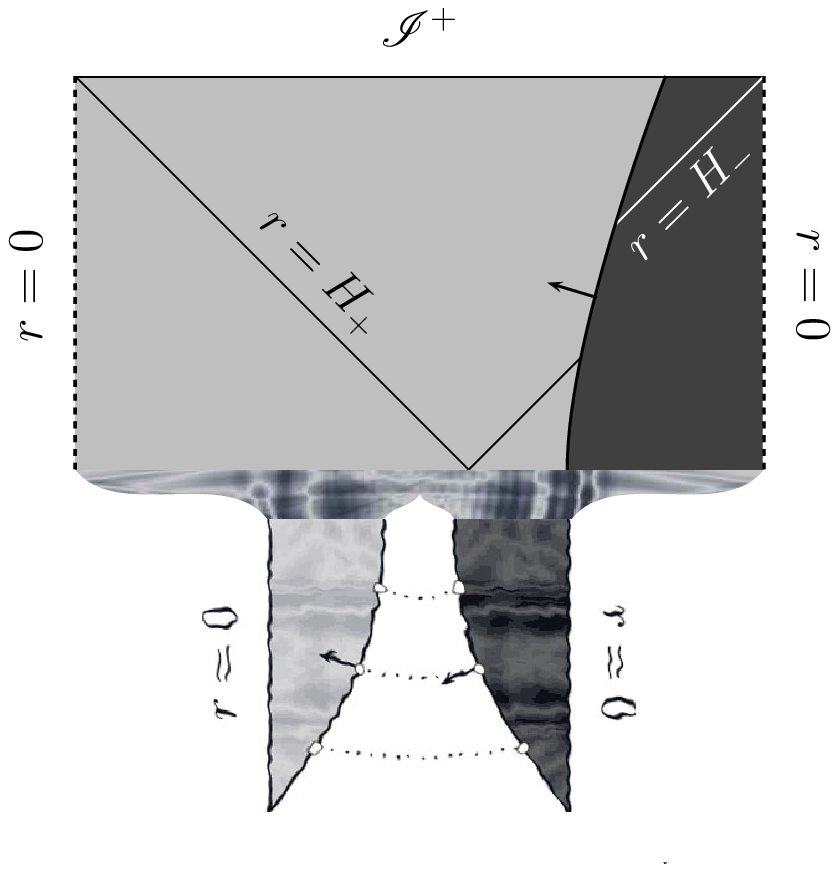}
\cr
$\mathrm{dS} _{+}/\mathrm{dS} _{-}$ & Tunnelling
\cr
\hline
\end{tabular}
\caption{\label{fig:junC_3}Case C3: dS/dS, $\epsilon_{-}=\epsilon_{+}=+1$}
\end{center}
\end{figure}

The last junction of the C type is C3, for which the usual set of diagrams is shown in
figure \ref{fig:junC_3}. As all C type ones, this is a junction between two de Sitter
spacetimes and since $\beta < -1$ the relation between the cosmological constants
is $\lambda _{+} > \lambda _{-}$, so that $H _{+} < H _{-}$. At this stage we might
wonder if this junction is the same as C1 (mirroring what happens between the junctions
of the A1 and A3 type). We will see that this is indeed the case. The normals now
point in both spacetimes toward the direction of increasing radii (since
$\epsilon _{\pm} = +1$). Thus the junction is as in the bottom left part of
figure \ref{fig:junC_3}. When we consider also the configuration before tunnelling, we see that the diagram in
the bottom right part of figure \ref{fig:junC_1} is the
``switched color and reflected'' version of the one in the bottom right part of
figure \ref{fig:junC_3}: we remember now that the in the C1 case $\lambda _{-}$ in
the light gray part was bigger than $\lambda _{+}$ in the dark side, but now exactly the
opposite happens. Thus, despite the different coloring, case C3 represents the same physical
situation of case C1. This observation makes interesting to complete the analysis for the
remaining two cases, which is done in the following subsection.

\subsection{Type D}

To conclude the analysis of the global spacetime structure we have to analyze
what happens for the last kind of junctions, those of type $D$. These are junctions
between de Sitter and anti--de Sitter spacetimes, so that the pretunnelling configuration will
also change accordingly, as we have discussed for the previous
cases.

\begin{figure}[!ht]
\begin{center}
\begin{tabular}{|c|c|}
\hline
\includegraphics[height=4cm]{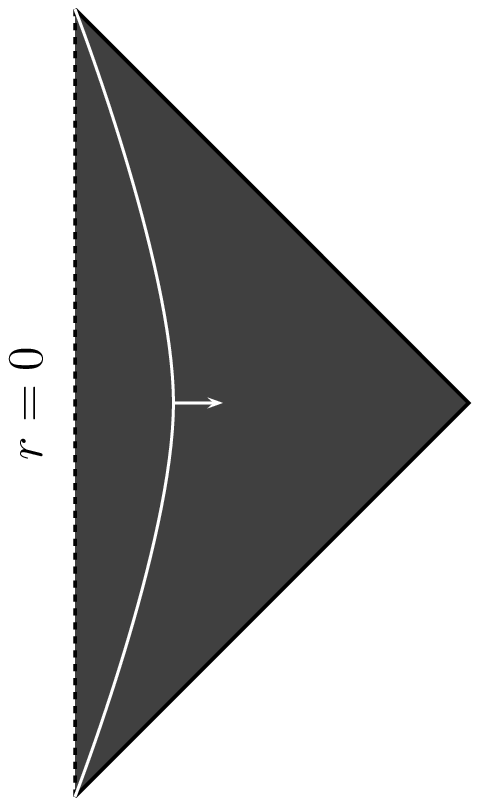}
&
\includegraphics[height=4cm]{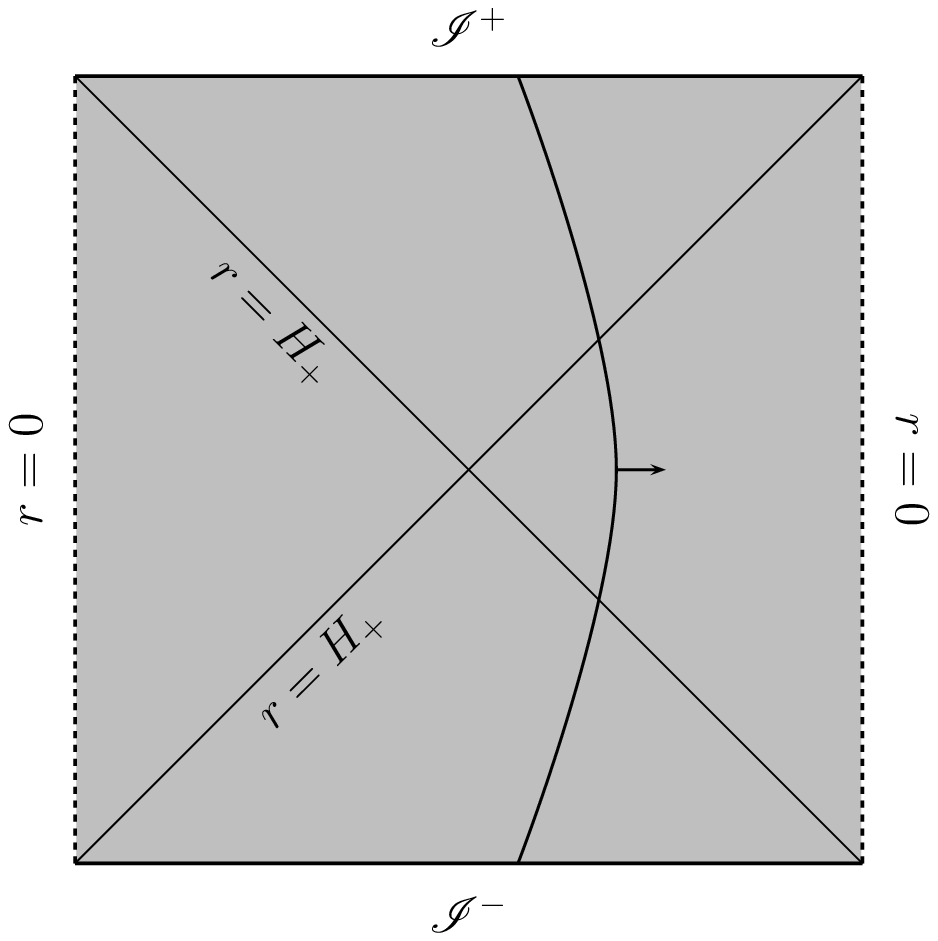}
\cr $\mathrm{AdS} _{-} , \epsilon _{-}=+1$ & $\mathrm{dS} _{+} , \epsilon _{+}=-1$ \cr
\hline
\includegraphics[height=4cm]{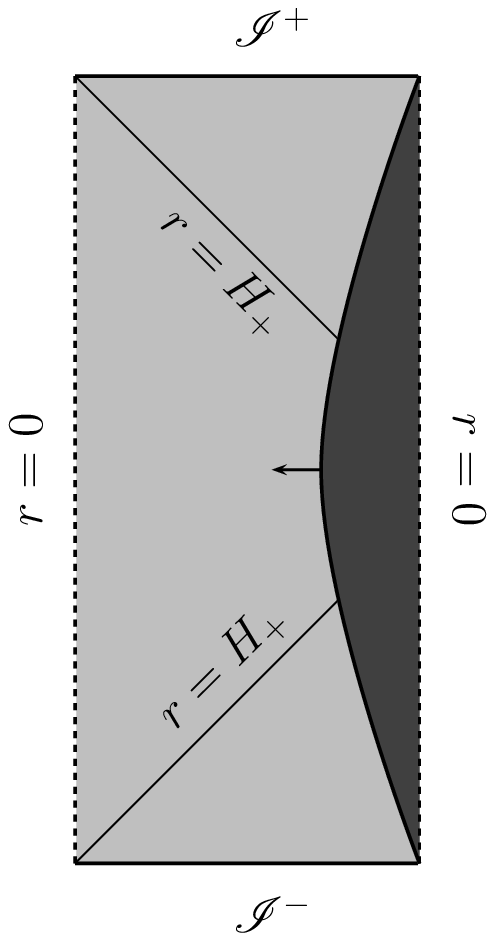}
&
\includegraphics[height=4cm]{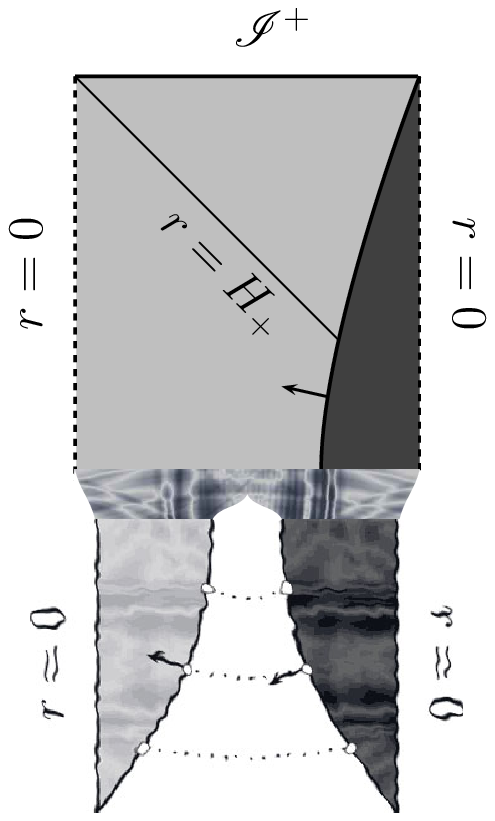}
\cr
$\mathrm{dS} _{+}/\mathrm{AdS} _{-}$ & Tunnelling
\cr
\hline
\end{tabular}
\caption{\label{fig:junD_2}Case D2: AdS/dS, $\epsilon _{-} = +1$;$\epsilon _{+} = -1$}
\end{center}
\end{figure}

The case of the junction of type $D2$ is characterized by the following signs:
$\epsilon _{\pm} = \mp 1$. This means that in the anti--de Sitter side the normal points
in the direction of increasing radii, whereas in the de Sitter part, it points in the
direction of decreasing radii. Thus the junction, shown in the bottom left corner
of figure \ref{fig:junD_2}, is the mirror image of the B2 case. Again, the considerations
in section \ref{sec:tunspatimstr} determine the geometry before the
tunnelling. This brings, in complete analogy with
the B2 case, to the picture of the tunnelling process given in the bottom right
part of figure \ref{fig:junD_2}; again, apart from the different colors, this case
is the same as B2 also in connection with the semiclassical tunnelling process.

\begin{figure}[!ht]
\begin{center}
\begin{tabular}{|c|c|}
\hline
\includegraphics[height=4cm]{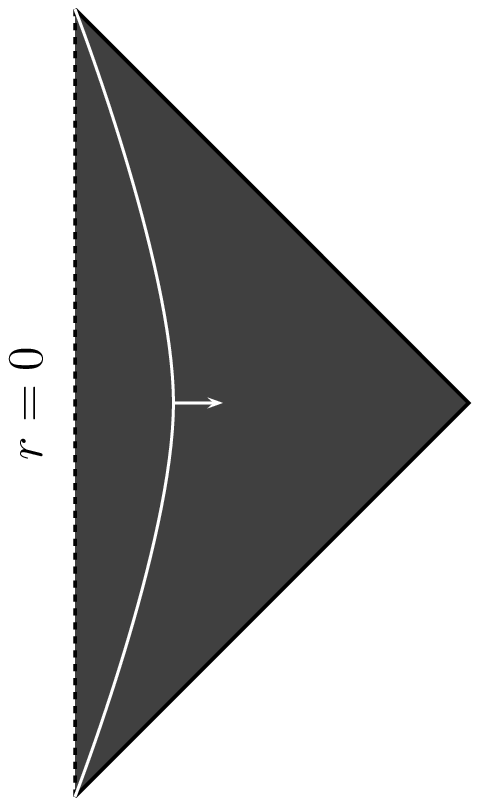}
&
\includegraphics[height=4cm]{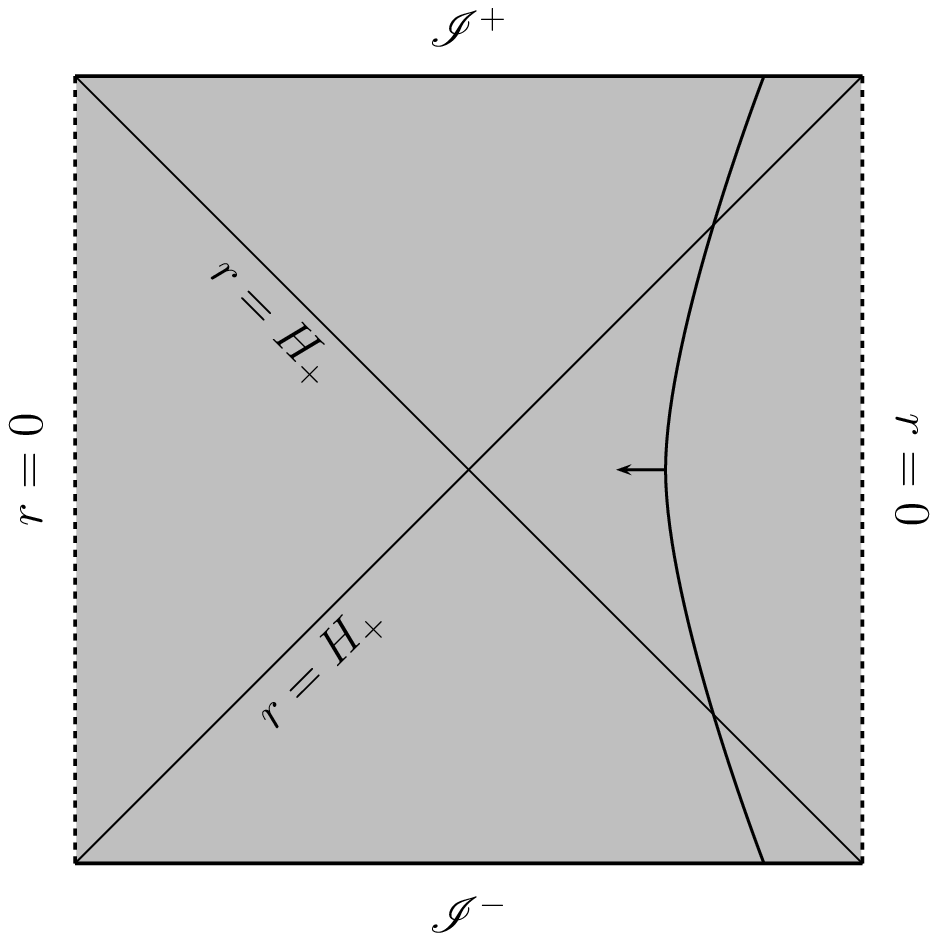}
\cr $\mathrm{AdS} _{-} , \epsilon _{-}=+1$ & $\mathrm{dS} _{+} , \epsilon _{+}=+1$ \cr
\hline
\includegraphics[height=4cm]{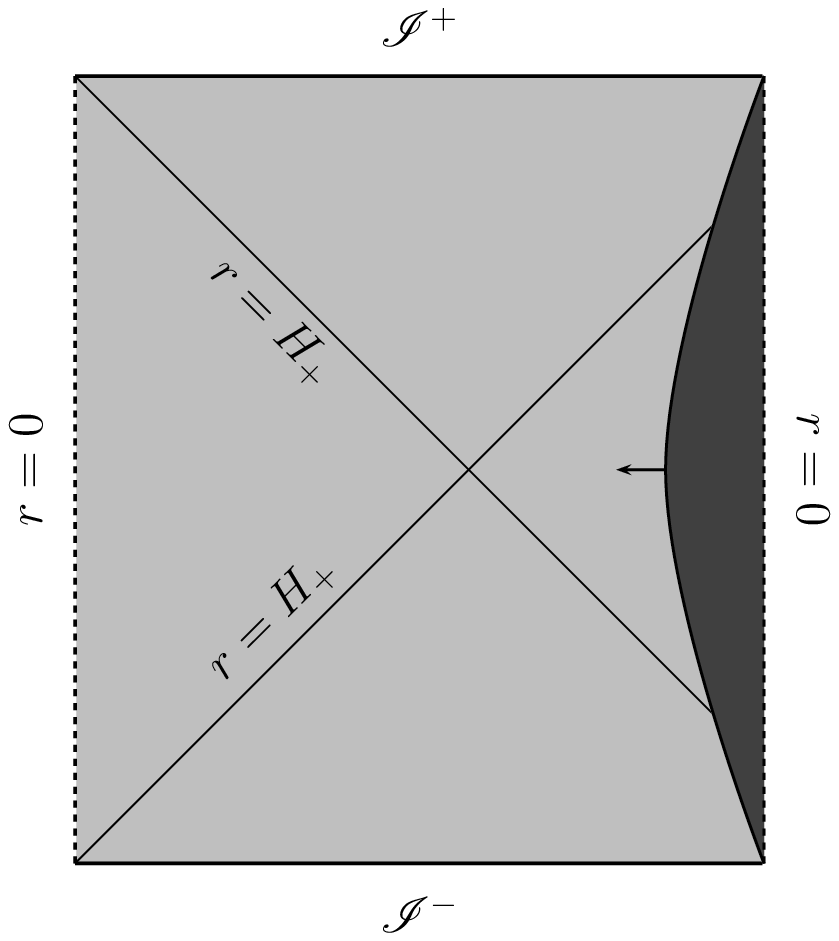}
&
\includegraphics[height=4cm]{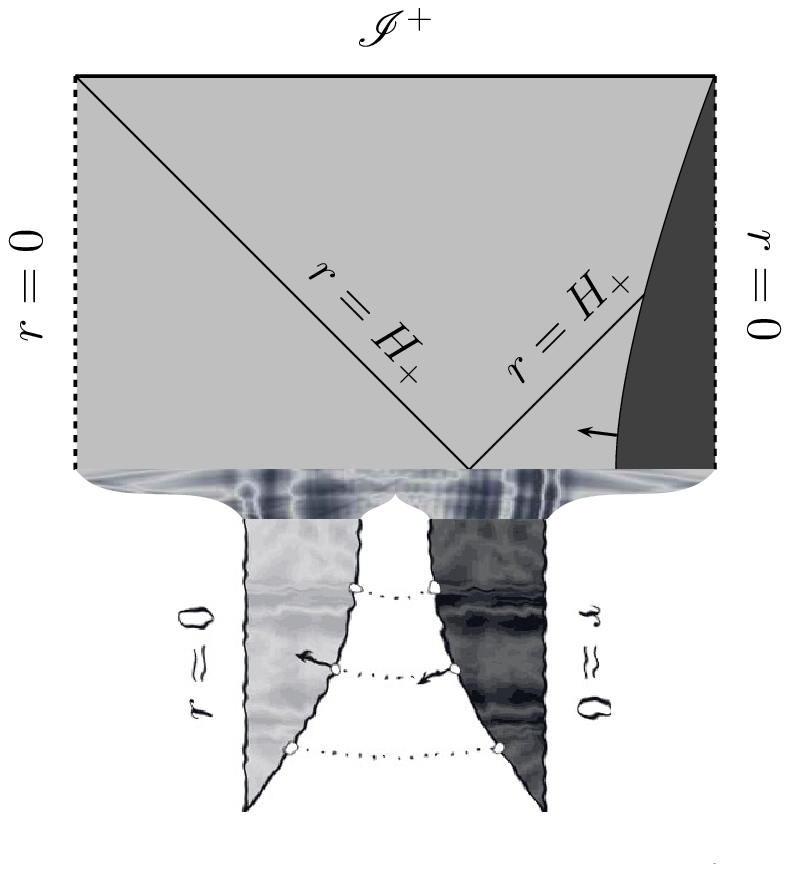}
\cr
$\mathrm{dS} _{+}/\mathrm{AdS} _{-}$ & Tunnelling
\cr
\hline
\end{tabular}
\caption{\label{fig:junD_3}Case D3: AdS/dS, $\epsilon _{-} = \epsilon _{+} = +1$}
\end{center}
\end{figure}

We have thus only one case left, which is quickly dealt with, namely D3. It is now not difficult
to anticipate that this process will turn out to be identical to B1, that appears in figure \ref{fig:junB_1}.
Indeed the signs are now $\epsilon _{\pm} = +1$, so that in both spacetime the normals point in
the direction of increasing radii. By the usual procedure we are thus led to the spacetime
diagram shown in the bottom left part of figure \ref{fig:junD_3}. Part of this diagram will constitute
the final state of the tunnelling process. The initial state is obtained as usual and the picture of the
tunnelling process is as in the bottom right part of figure \ref{fig:junD_3}.

\subsection{Comments}

From the analysis developed so far we have seen that, although in principle we have
ten different tunnelling processes, in fact process A1 is the same as A3, process B1
is the same as D3, process B2 is the same as A2 and process C1 is the same as C3.
We are thus left with only six distinct processes. From ``symmetry consideration'',
it is natural to notice that $| \beta |$, not $\beta$ itself, is the relevant
quantity (together with $\alpha$) to classify the possible tunnelling configurations.
We will thus not be surprised if the tunnelling amplitude will be a function of
$| \beta |$, or, which is the same, an \emph{even function of $\beta$}. We would also like to note
already here that all the tunnelling processes can be interpreted as a very large expansion
of a mall \emph{quantum} region of spacetime.


\end{document}